\documentclass[debug]{rmaa}

\usepackage{graphicx}
\usepackage{xspace}
\usepackage{txfonts}
\usepackage{multirow}
\usepackage{lscape}
\usepackage{natbib,twoopt}
\usepackage{color}% Nueva línea de texto. 
\usepackage{soul}
\usepackage{longtable}
\usepackage{geometry}
\usepackage[breaklinks=true]{hyperref} % to avoid \citeads line fills
\bibpunct{(}{)}{;}{a}{}{,} % to follow the A&A style

\makeatletter
\newcommandtwoopt{\citeads}[3][][]{\href{http://adsabs.harvard.edu/abs/#3}%
{\def\hyper@linkstart##1##2{}%
\let\hyper@linkend\@empty\citealp[#1][#2]{#3}}}
\newcommandtwoopt{\citepads}[3][][]{\href{http://adsabs.harvard.edu/abs/#3}%
{\def\hyper@linkstart##1##2{}%
\let\hyper@linkend\@empty\citep[#1][#2]{#3}}}
\newcommandtwoopt{\citetads}[3][][]{\href{http://adsabs.harvard.edu/abs/#3}%
{\def\hyper@linkstart##1##2{}%
\let\hyper@linkend\@empty\citet[#1][#2]{#3}}}
\newcommandtwoopt{\citeyearads}[3][][]%
{\href{http://adsabs.harvard.edu/abs/#3}
{\def\hyper@linkstart##1##2{}%
\let\hyper@linkend\@empty\citeyear[#1][#2]{#3}}}
\makeatother

\newcommand{\ergs}{\ensuremath{\mathrm{erg\,s}^{-1}}\xspace}

\newcommand{\ergscm}{\ensuremath{\mathrm{erg\,s}^{-1}\mathrm{cm}^{-2}}\xspace}

\newcommand{\Msol}{\ensuremath{\mathrm{M}_{\odot}}\xspace}
\newcommand{\Rsol}{\ensuremath{\mathrm{R}_{\odot}}\xspace}

\newcommand{\xmm}{\textsl{XMM-Newton}\xspace}

\newcommand{\chandra}{\textsl{Chandra}\xspace}

\newcommand{\asca}{\textsl{ASCA}\xspace}
\newcommand{\maxi}{\textsl{MAXI}/GSC\xspace}

\newcommand{\centau}{Cen~X-3\xspace}

\usepackage{paralist}

\usepackage{psfrag,color}

\usepackage[latin1]{inputenc}

\title{Cen X--3 as seen by MAXI during six years} 

\author{
  \'A. Torregrosa,\altaffilmark{{2}} 
  J. J. Rodes-Roca,\altaffilmark{1,2}
J. M. Torrej\'on,\altaffilmark{1,2}
G. Sanjurjo-Ferr\'in\altaffilmark{{2}}
  and G. Bernab\'eu\altaffilmark{1,2}}

\altaffiltext{1}{Department of Physics, Systems Engineering and Signal Theory, University of Alicante,
              03080 Alicante, Spain.}

\altaffiltext{2}{University Institute of Physics Applied to Sciences and Technologies,
              University of Alicante, 03080 Alicante, Spain.}

\shortauthor{\'A. Torregrosa et al.}
\shorttitle{Cen X-3 as seen by MAXI during six years.}

\fulladdresses{
\item \'A. Torregrosa, J. J. Rodes-Roca, J. M. Torrej\'on, G. Sanjurjo-Ferr\'in and G. Bernab\'eu: University Institute of Physics Applied to Sciences and Technologies, University of Alicante, 03080 Alicante, Spain (aat34@alu.ua.es).
\item J. J. Rodes-Roca, J. M. Torrej\'on and G. Bernab\'eu: Department of Physics, Systems Engineering and Signal Theory, University of Alicante, 03080 Alicante, Spain.}

\listofauthors{\'A. Torregrosa, J. J. Rodes-Roca, J. M. Torrej\'on, G. Sanjurjo-Ferr\'in, G. Bernab\'eu}
\indexauthor{Torregrosa, \'A.}
\indexauthor{Rodes-Roca, J. J.}
\indexauthor{Torrej\'on, J. M.}
\indexauthor{Sanjurjo-Ferr\'in, G.}
\indexauthor{Bernab\'eu, G.}

\abstract{The aim of this work is to study both light curve and orbital phase spectroscopy of this source taking advantage of the \maxi observation strategy. We have investigated the spectral and light curve properties of the X-ray emission from \centau along the binary orbit. These studies allow delimiting the stellar wind properties and its interactions with the compact object. A timing
analysis of light curves in different energy bands was carried out. From the analysis of the light curve, we have estimated the orbital period of the binary system and also found possible QPOs around a superorbital period of $P_\mathrm{superorb} = 220\pm5$ days. Both orbital phase-averaged and phase-resolved spectra were extracted and analysed in the 2.0--20.0 keV energy range. We have defined and compared the high and low states spectra with the averaged spectrum. Two models have described spectra satisfactorily, a partial absorbed Comptonization of cool photons on hot electrons plus a power law and a partial absorbed blackbody plus a power law, both modified by adding Gaussian lines. The radius of the blackbody emitting area has been determined and the high value of the X-ray luminosity in the averaged spectrum indicates that the accretion mode is not only due to the stellar wind.}

\resumen{Se pretende estudiar la curva de luz y la espectroscop\'ia en fase orbital de esta fuente gracias a la estrategia de observaci\'on de \maxi. Hemos investigado las propiedades espectrales y de la curva de luz en la emisi\'on de rayos X de \centau a lo largo de su \'orbita. Estos estudios permiten delimitar las propiedades del viento estelar y sus interacciones con el objeto compacto. Un an\'alisis temporal de las curvas de luz en distintas bandas energ\'eticas ha sido realizado. Por medio de este an\'alisis hemos estimado el periodo orbital del sistema y hemos encontrado posibles QPOs alrededor de un superperiodo orbital de $P_\mathrm{superorb} = 220\pm5$ d\'ias. Hemos extra\'ido y analizado los espectros medio y en fase orbital en el rango 2.0--20.0 keV por medio de dos modelos: una comptomizaci\'on parcialmente absorbida de fotones  fr\'ios en electrones calientes m\'as una ley de potencias y un cuerpo negro parcialmente absorbido m\'as una ley de potencias, ambos modificados por l\'ineas gaussianas. Hemos definido y comparado los espectros de los estados altos y bajos con el espectro medio. Hemos determinado el radio de la region de emisi\'on y el alto valor de la luminosidad del espectro medio indica un modo de acreci\'on no debido solamente al viento estelar.}

\addkeyword{pulsars: individual: Cen X--3}
\addkeyword{stars: supergiants}
\addkeyword{X-rays: binaries}

\begin{document}
% Typeset article header
\maketitle

\section{Introduction}
\label{introduction}
\centau is an eclipsing high-mass X-ray binary system formed by an O-type donor star and a neutron star. Its complex X-ray behaviour makes both its temporal and spectral analysis extremely important to better understand its properties. So far, it is the only high-mass X-ray binary (HMXB) system in the Milky Way where mass transfer onto a neutron star occurs directly from the surface of the donor star (via Roche lobe overflow). The source has been discovered in 1967 \citep{1967PhRvL..19..681C} % chodil et al.
and \citet{1974ApJ...192L.135K} %Krzeminski
estimated the distance as $\sim$ 8 kpc. The distance to Cen X-3 obtained from the European Space Agency (ESA) mission Gaia\footnote{https://www.cosmos.esa.int/gaia} Early Data Realease 3 (GEDR3) is $d (kpc) = 6.8^{+0.6}_{-0.5}$. It has been derived by using the parallax measure, the source's G-band magnitude and BP-RP colour \citep{2021AJ....161..147B}, %Bailer-Jones et al. (2021)
called by them ''photogeometric'' distance.

It consists of a neutron star (NS) and a giant star O6--8 III companion called V779 Cen \citep{1979ApJ...229.1079H} with mass  $\sim$20 \Msol \citep{2007A&A...473..523V} %van der Meer et al 2007 
and radius $\sim$12 \Rsol \citep{2011ApJ...737...79N}. % naik et al. 2011
\citet{2011ApJ...730...25R} %Rawls et al. (2011) 
calculated the NS mass by using a Monte Carlo method and assuming a Roche lobe filling factor between 0.9 and 1.0, $M_\mathrm{NS}$ = (1.35$\pm$0.15) \Msol. However, applying another technique based on eclipsing light curve analysis and including an accretion disc around the NS, they derived a final mass for this system of (1.49$\pm$0.08) \Msol.
The binary orbit is almost circular, eccentricity $e < 0.0016$ \citep{1997ApJS..113..367B}, % bildsten et al. 1997
with an orbital period of $\sim$2.1 days, determined from regular X-ray eclipses \citep{1972ApJ...172L..79S}. %Schreier et al. 1972

The observed average high X-ray emission $\sim 10^{37}$ \ergs compared to that observed in wind-fed accreting systems $\sim 10^{36}$ \ergs \citep{2017SSRv..212...59M, 2019NewAR..8601546K} suggests additional structures
in the NS environment. Studies of the light curve of \centau
\citep{1986A&A...154...77T} % tjemkes et al. 1986
and the observed overall NS spin-up trend of 1.135 ms yr$^{-1}$ with fluctuations on time scales of years \citep{1996ApJ...456..316T} % tsunemi et al. 1996
together with the detection of quasi periodic oscillations (QPOs) from the source \citep{1991PASJ...43L..43T, 2008ApJ...685.1109R} % takeshima et al. 1991, raichur & paul 2008
showed evidences of an accretion disk, due to Roche lobe overflow. Other structures such as additional gas stream or accretion wake might also be present in the accretion scenario in this source \citep{1988MNRAS.232..199S, 2008ApJ...675.1487S}. % stevens, i. r. 1988, suchy et al. 2008

Thus, variability of model parameters along the orbital phase allow us not only to investigate the characteristics of the circumstellar matter around the NS but also to trace permanent wind structures. Many authors have reported on the iron emission lines of \centau \citep{1996PASJ...48..425E, 2005ApJ...634L.161I, 2012BASI...40..503N, 2017symm.conf..159R, 2019ApJS..243...29A}
%Ebisawa et al., Iaria et al., naik and paul, Rodes et al., aftab et al.
around the orbital period. Although \emph{MAXI} was able to detect the Fe K$\alpha$, it cannot resolve it properly. However, changes in the central energy of the iron line
with the orbital phase pointed to a coexistence of two iron lines at different energies \citep{2017symm.conf..159R}. % rodes et al. 2017
Therefore, by using \asca description of the iron emission lines of \centau we fitted three Gaussian profiles to the \maxi data fixing some of the line parameters to estimate their equivalent widths.

In this paper we present the spectroscopic and light curve analysis of \centau observed with \emph{MAXI}\@. \maxi data covering the entire orbit and extending over more than six years. Orbital phase-averaged and phase-resolved spectroscopy were performed applying several models in the 2--20 keV energy range. Observations and data reduction are described in \S~\ref{data}, timing analysis is presented in \S~\ref{timing}, orbital phase-averaged spectrum and orbital phase-resolved spectra results are discussed in \S~\ref{spectra}, and \S~\ref{conclusion} contains the summary of the main findings.

\section{Observations and data}
\label{data}

\emph{MAXI} is an X-ray monitor on board the International Space Station (ISS) since August 2009 \citep[][]{2009PASJ...61..999M}. %Matsuoka et al. 2009
Every $\sim$92 minutes, it scans almost the entire sky in each ISS orbit, observing a particular source for about 40--150 s \citep{2011PASJ...63S.635S}, %sugizaki et al. 2011
depending on the position of the object. It consists of two types of X-ray slit cameras, the Gas Slit Camera \citep[GSC,][]{2011PASJ...63S.623M} in the 2.0--20.0 keV, %Mihara et al. 2011  
and the Solid-state Slit Camera \citep[SSC,][]{2011PASJ...63..397T} %Tomida et al. 2011
operating in the 0.7--7.0 keV energy range. The in-orbit performance of GSC and SSC is presented in \citet{2011PASJ...63S.635S} and \citet{2010PASJ...62.1371T}, %tsunemi et al. 2010
respectively.

\section{Timing analysis}
\label{timing}

Firstly, we have extracted the \maxi on-demand light curves of \centau with a time resolution of one ISS orbit in \textbf{five} energy bands, 2.0--20.0, 2.0--4.0, 4.0--10.0, 10.0--20.0 and
5.7--7.5 keV\@. To analyse the light curves we used \emph{Python} and the \emph{Starlink} software package\footnote{\href{http://starlink.eao.hawaii.edu/starlink}{\url{http://starlink.eao.hawaii.edu/starlink}}}.

Secondly, we have used the Lomb-Scargle 
technique \citep{1989ApJ...338..277P} %Press & Rybicki
to determine the orbital period from the original light curve 2.0--20.0 keV and have obtained $P_\mathrm{orb}= 2.0870\pm0.0005$ days (Figure~\ref{figure_temporal_2}),
which is comparable to that obtained by \citet{1992ApJ...396..147N}, % nagase et al. 1992
\citet{2010MNRAS.401.1532R}, % raichur and paul 2010
\citet{2015A&A...577A.130F} % falanga et al. 2015
or \citet{2017symm.conf..159R}. The errors in the periods were estimated using the \emph{Peaks} tool inside the time-series analysis package \emph{Period} in \emph{Starlink}.  The effect of the barycentric correction on the light curves was found to be negligible and did not need to be taken into consideration. Then, we folded the light curves with the best orbital period to produce energy-resolved orbital intensity profiles using the ephemeris from \citet{2015A&A...577A.130F}. % falanga et al. 2015.
In Table~\ref{tab:99} we compiled the ephemeris data used for timing calculations.% falanga et al. 2015.

\begin{table}[h!tb]%[scale=2.0,width=2\columnwidth]
 \caption{Ephemeris data used for timing calculations}
    \centering
    \begin{tabular}{lccc} \toprule\toprule     
$T_{0,ecl}$ (MJD) & $50506.788423\pm 0.000007$ \\
$P_{\mathrm{orb}}$ (d)  & $2.08704106 \pm 0.00000003$ (Falanga et al. (2015)) \\
$\dot{P}_{\mathrm{orb}}/P_{\mathrm{orb}}$ (10$^{-6}$ yr$^{-1}$)  & $-1.800 \pm 0.001$  \\
     \bottomrule
    \end{tabular}
    
    \label{tab:99}
\end{table}

\begin{figure}[hbtp]
  \centering
  \includegraphics[angle=0.0,width=\columnwidth]{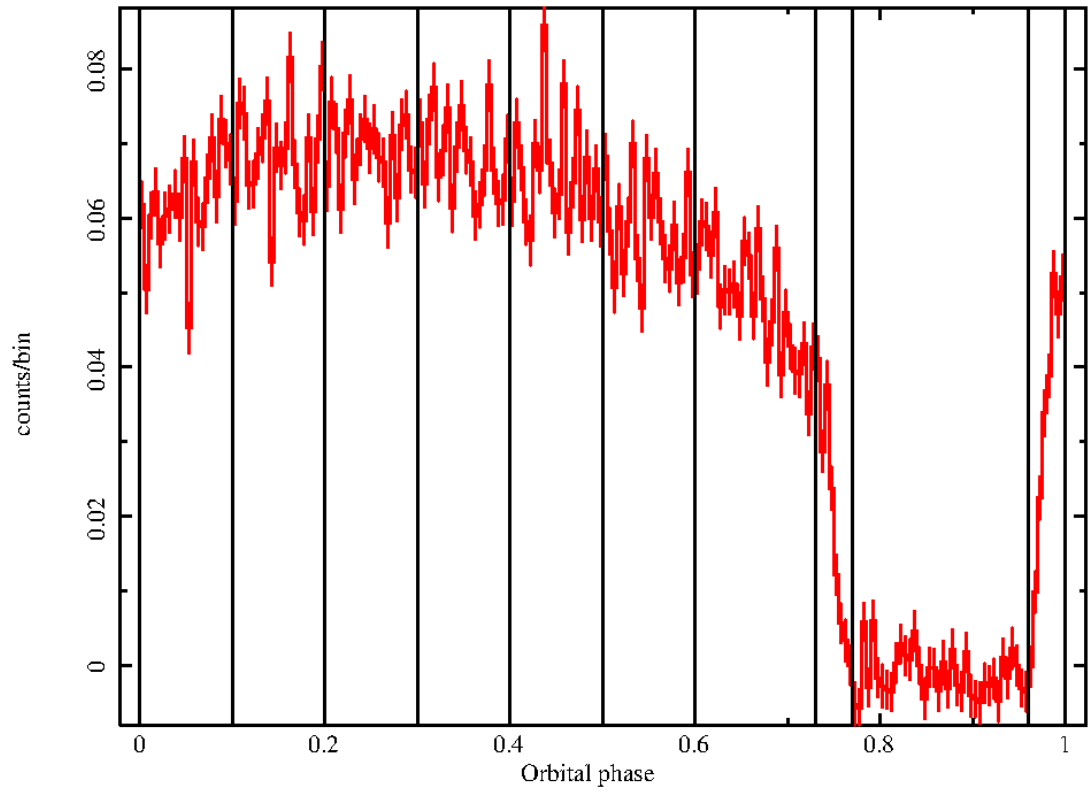}  
  \caption[]{\label{figure_temporal_1} %
    Background subtracted light curve in 10.0--20.0 keV energy range.   
  }
\end{figure}

\begin{figure}[hbtp]
  \centering
  \includegraphics[angle=-90.0,width=\columnwidth]{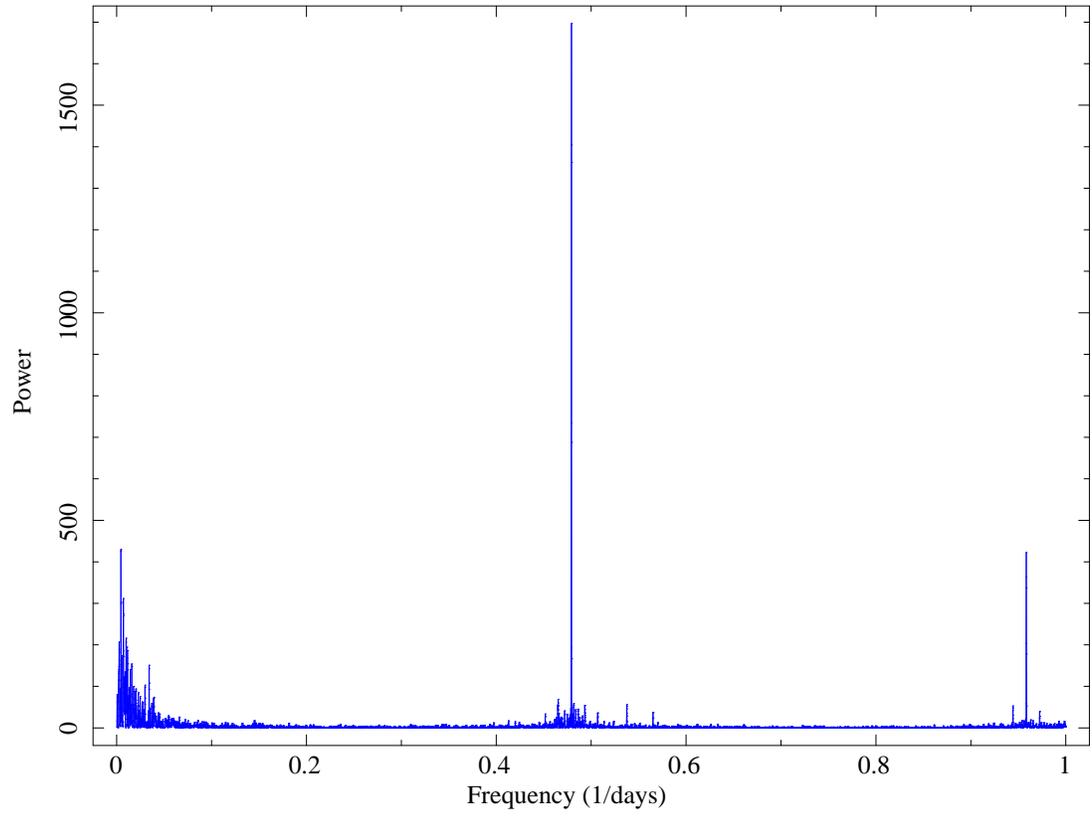}  
  \caption[]{\label{figure_temporal_2} %
   Lomb-Scargle periodogram of the 2.0--20.0 keV light curve.   
  }
\end{figure}

As a sample, the resulting orbital light curve
between 10.0--20.0 keV is plotted in Figure~\ref{figure_temporal_1} where the strongest changes are observed
as the NS ingresses and egresses from eclipse.
We defined ten orbital phase bins corresponding to phase intervals [0.0--0.1] (post-egress), [0.1--0.2], [0.2--0.3], [0.3--0.4], [0.4--0.5], [0.5--0.6], [0.6--0.73] (pre-ingress),
[0.73--0.77] (ingress), [0.77--0.96] (which corresponds to the total eclipse) and [0.96--1.0] (egress).

\begin{figure}[hbtp]
  \centering
  \includegraphics[angle=-90.0,width=\columnwidth]{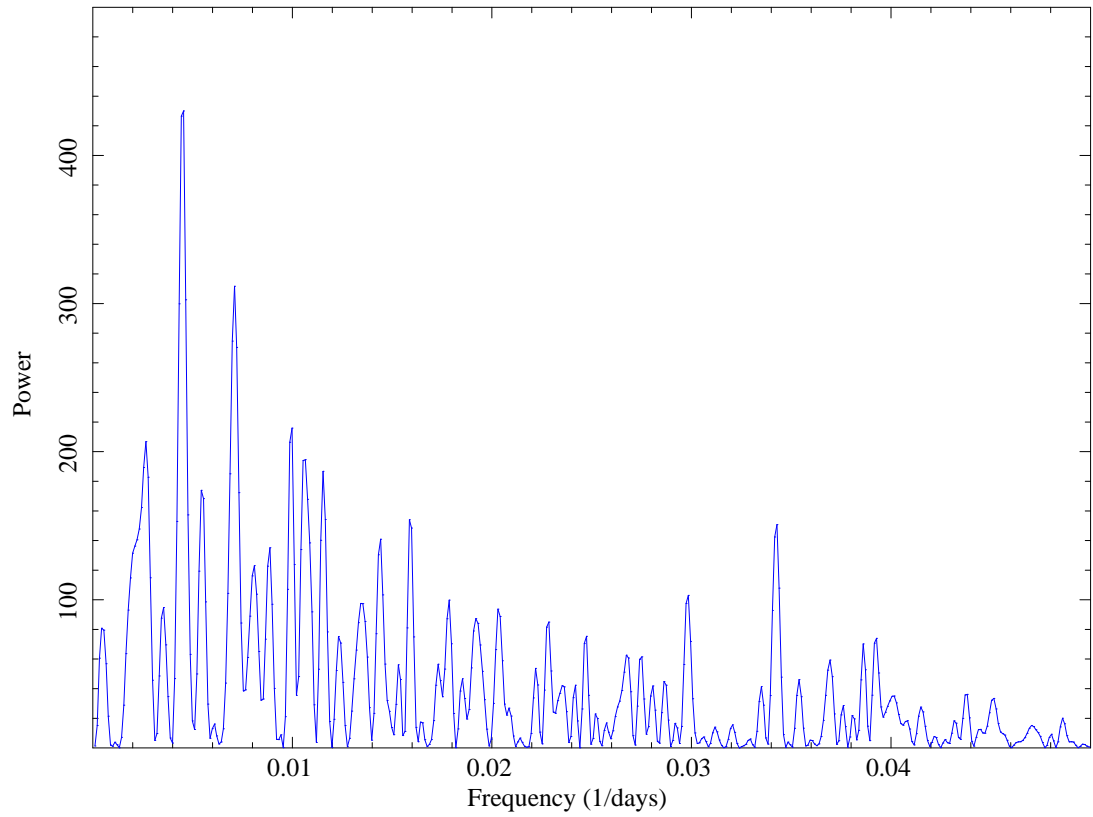}  
  \caption[]{\label{figure_temporal_3} %
     Zoom of the QPOs in the original 2.0--20.0 keV light curve.   
  }
\end{figure}

Another two maximum peaks were present in the Lomb-Scargle periodogram. The first one with a power of 430.03 corresponds to a period of $220\pm5$ days, which might be consistent with the superorbital period of 93.3--435.1 days reported by \citet{2014JPSCP...1a3104S}. %Sugimoto et al.   
The second one with a power of 422.48 corresponds to a period of $\sim$ 1.04 days, which is a harmonic of the orbital period. 
In addition, around the superorbital period, there are a few more peaks that can be interpreted as QPOs (Figure~\ref{figure_temporal_3}), as also pointed out by
\citet{2008MNRAS.387..439R}, \citet{2008ApJ...685.1109R},  
\citet{1991PASJ...43L..43T} %Takeshima et al.
and \citet{1983ApJ...273..709P}. %Priedhorsky et al.

\begin{figure}[hbtp]
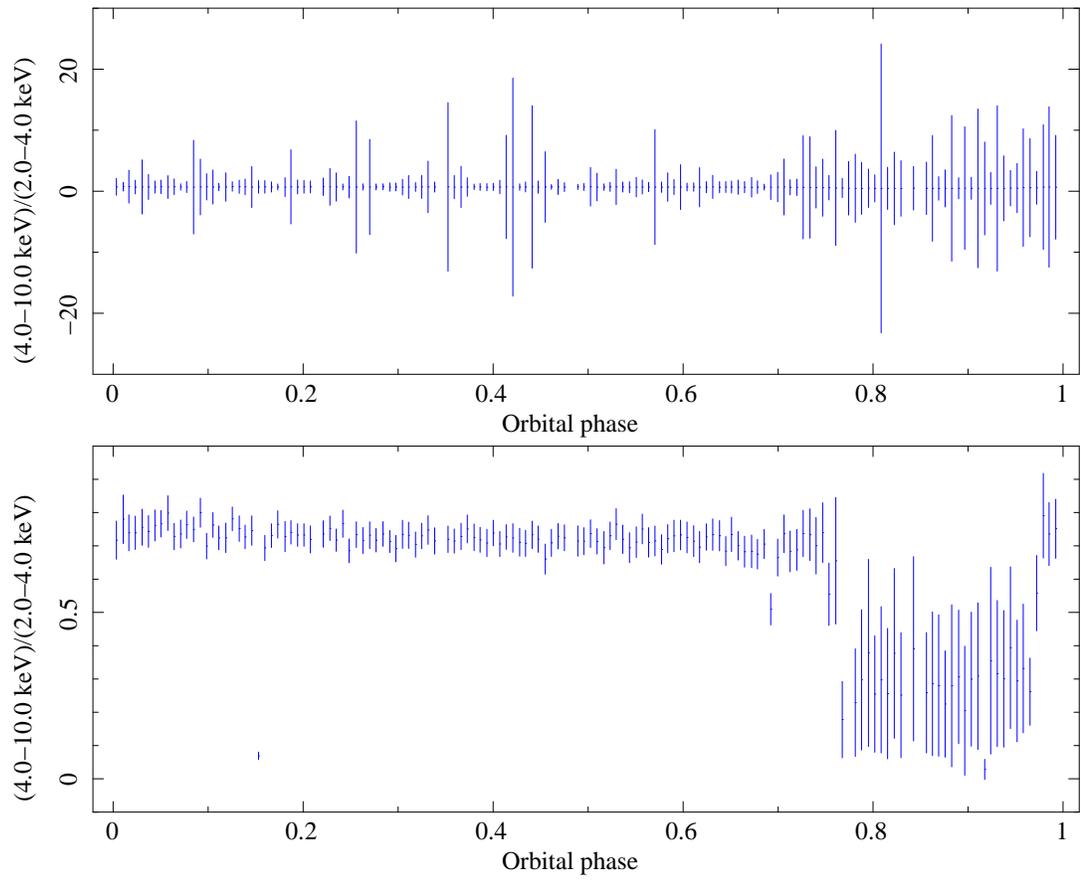

  \centering
  \includegraphics[angle=-90.0,width=\columnwidth]{cr_low_med_v2.ps}
  \includegraphics[angle=-90.0,width=\columnwidth]{cr_pond_low_med_v2.ps} 
  \caption[]{\label{figure80} % 
    Hardness ratio $H/S = (4.0-10.0\, \, \mathrm{keV})/(2.0-4.0\, \, \mathrm{keV})$. \emph{Top panel}: using simple average. \emph{Bottom panel}: using weighted average.
  }
\end{figure}

It is expected that HMXBs show strong absorption at low energies (below 4 keV), complex iron emission lines between 6.4 keV and 7.2 keV and an energy cutoff at high energies (greater than 10 keV). In order to analyse the light curves variability the \maxi energy range (2.0--20.0 keV) has been divided into 2.0--4.0 keV (low energy), 4.0--10.0 keV (medium energy), 5.7--7.5 keV (iron complex emission lines) and 10.0--20.0 keV (high energy) bands.
Therefore, we have calculated the hardness ratio defined as $H/S$, where $H$ are the net counts in the hard band and $S$ are those obtained in the soft band, between the light curves: (5.7--7.5 keV)/(2.0--4.0 keV), (4.0--10.0 keV)/(5.7--7.5 keV), (10.0--20.0 keV)/(5.7--7.5 keV), (10.0--20.0 keV)/(2.0--4.0 keV), (10.0--20.0 keV)/(4.0--10.0 keV) and (4.0--10.0 keV)/(2.0--4.0 keV). From the folded light curve, the hardness ratio measurements obtained for each pointing along the orbit had large error bars. Therefore, they were averaged over 150 points and the uncertainties were estimated using error propagation. The overall profile shape was quite similar and consistent with a constant value, so we could not identify any morphology or tendency (see Figure~\ref{figure80}, top panel). Figure~\ref{figure80}, bottom panel, shows the hardness ratio using a weighted average over 150 bins. The hardness ratio is consistent with a constant ($HR \sim 0.75$) for the out of eclipse indicating there is no significant change in the spectral shape. On the other hand, during the eclipse the brightness of \centau is also consistent with a constant value ($HR \sim 0.3$) but 2.5 times lower than out of eclipse. Both the drop in brightness at eclipse ingress and the rise in brightness at eclipse egress are quite sharp.

\begin{figure}[hbtp]
  \centering
  \includegraphics[angle=-90.0,width=\columnwidth]{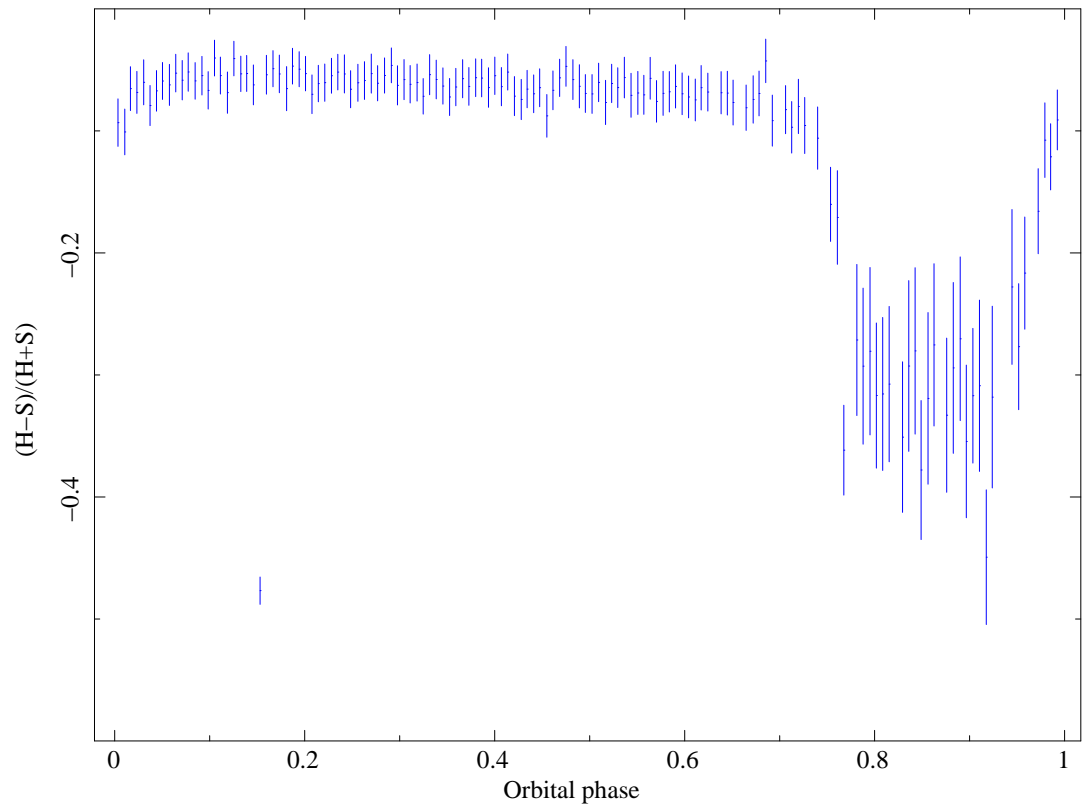} 
  \caption[]{\label{figure32} %
    Hardness curve $(H-S)/(H+S)$ using weighted average, where $H = 4.0-10.0$ keV and $S = 2.0-4.0$ keV energy bands.}
\end{figure}

The process to obtain the hardness curves $(H-S)/(H+S)$ is the same used to calculated $H/S$, where $H$ = 4.0--10.0 keV and $S$ = 2.0--4.0 keV (Figure~\ref{figure32}). The general trend of the hardness curve is very similar to that of the hardness ratio, although the decrease and increase before and after the eclipse is smoother. Moreover, as can be seen in Figures~\ref{figure5} (equation~(\ref{ec1})) and \ref{figure6} (equation~\ref{ec4})) it is consistent with the behaviour of the unabsorbed flux variation.

\begin{figure}[hbtp]
  \centering
  \includegraphics[angle=-90.0,width=\columnwidth]{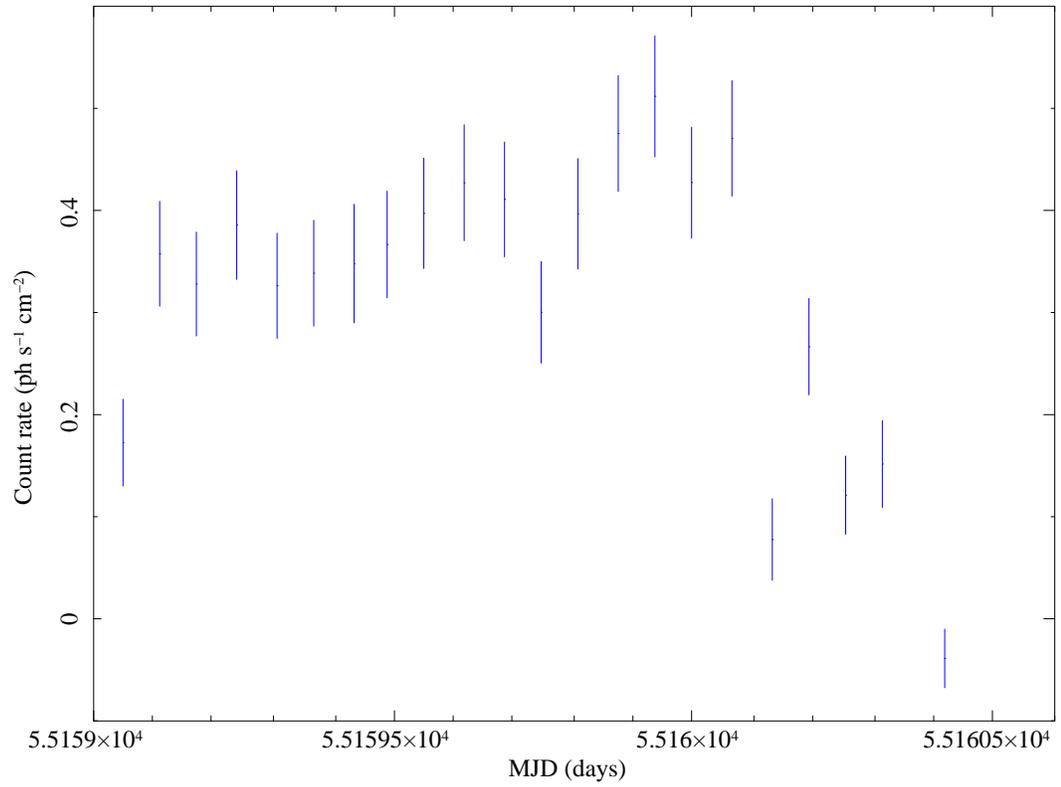} 
  \caption[]{\label{figure50} %
    Light curve in 2.0--20.0 keV energy range for an orbital period of \centau.

  }
\end{figure}

During an orbital period of \centau, \maxi can perform up to 33 observations of the object (an example is shown in Figure~\ref{figure50}). However, for individual point X-ray sources, \maxi detector have very short exposures of about 60 s fifteen times a day which is not long enough to extract orbital phase-resolved spectra. As a consequence, it is necessary to accumulate observations in each orbital phase to obtain spectra with a good signal-to-noise ratio.


\section{Spectral analysis}
\label{spectra}

\subsection{Orbital phase averaged spectrum}
\label{averaged_spectrum}

\begin{figure}[hbtp]
  \centering
  \includegraphics[angle=-90.0,width=\columnwidth]{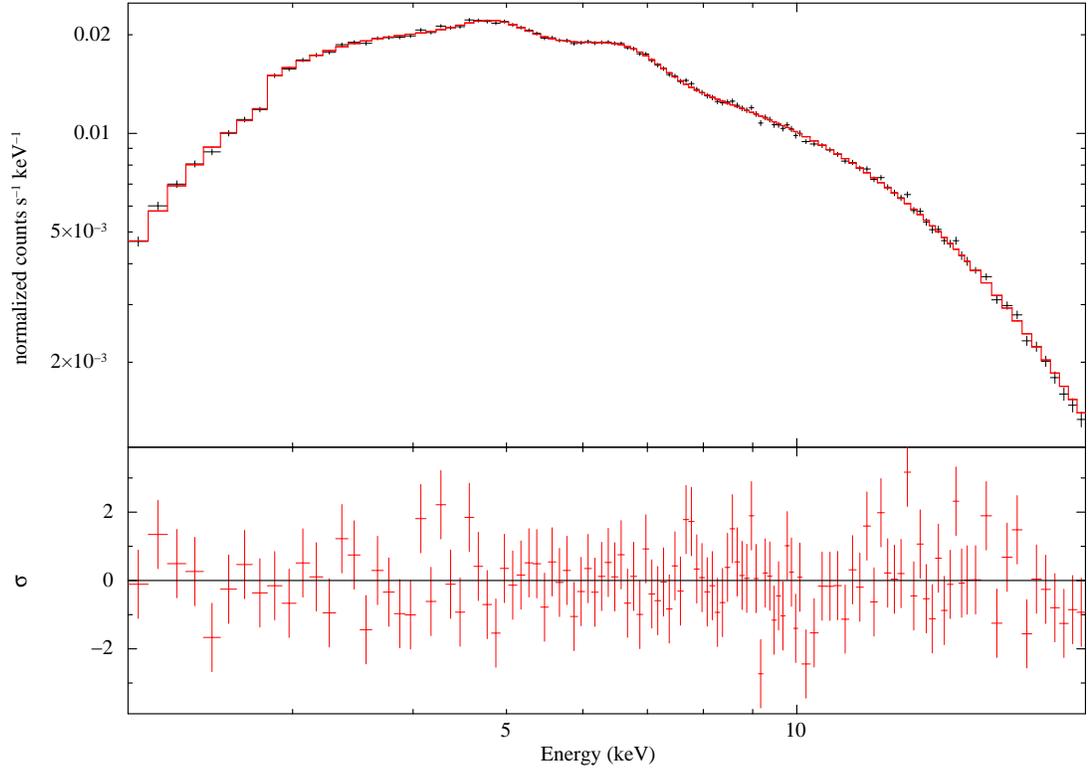}  
  \caption[]{\label{figure1} %
   Orbital phase averaged spectrum of \centau
    in the 2.0--20.0 keV band.
    \emph{Top panel}: data and 
     best-fit model described by equation~(\ref{ec1}).
    \emph{Bottom panel} shows the residuals between the spectrum and the model. The resulting fit parameters are reported in Table~\ref{tab:15}.
    
  }
\end{figure}

We have extracted the orbital phase averaged spectrum of \centau (see Figure~\ref{figure1}) with \maxi using the
\emph{MAXI} on-demand processing\footnote{\href{http://maxi.riken.jp/mxondem}{\url{http://maxi.riken.jp/mxondem}}}, carefully excluding any contamination by nearby brighter sources. For spectral analysis we used the \textsc{XSPEC} fitting package, released as a part of \textsc{XANADU} in the HEASoft tools. We tested both phenomenological and physical models commonly applied to accreting X-ray pulsars and rebinned all extracted spectra to obtain spectral bins by Gaussian distribution.

Absorbed comptonisation models have been successfully applied to HMXBs in the 2.0--20.0 keV energy range covered by \maxi, such as Vela X-1 \citep{2013A&A...554A..37D}, % doroshenko et al. 2013
4U 1538-52 \citep{2015A&A...580A.140R} and \centau \citep{2017symm.conf..159R}. We started to describe the orbital phase-averaged spectrum using a simple partial absorbed Comptonisation model modified by a Gaussian absorption line at $\sim$5 keV to compensate the incompleteness of the \maxi response (Nakahira private communication). An iron fluorescence emission line present in the spectrum is modelled with a Gaussian component, Fe K$\alpha$ at 6.4 keV. This model is described by equation~(\ref{ec11}).

\begin{equation}
  F(E) = \mathit{pcfabs} \times \mathit{gabs} \times \mathit{compST} + \mathit{Gaussian \,\, line}
  \label{ec11}
\end{equation}

where, in terms of \textsc{XSPEC}, \emph{pcfabs} is a partial covering fraction absorption that affects only a fraction $f$ of the model component multiplied by it, \emph{gabs} is the
Gaussian absorption line component, \emph{compST} is the
Comptonisation of cool photons on hot electrons \citep{1980A&A....86..121S}, and the Gaussian line is added to describe the Fe K$\alpha$ line. The absorption cross
sections were taken from \citet{1995A&AS..109..125V} % verner et al. 1996
and the abundances were set to those of \citet{2000ApJ...542..914W}. % wilms et al. 2000

Although this model describes the averaged spectra between 2.0--20.0 keV well ($\chi^{2}_{r}$ = 1.09), it cannot offer a good statistical and/or physical solution in all orbital phase-resolved spectra. For example, fitting the eclipse spectrum, i.e. when the direct X-ray emission is totally blocked by the companion, we obtained that the covering fraction factor was $f \approx$ 0. Model parameters were also not well constrained. As the aim is to satisfactorily describe both the averaged spectrum and the orbital phase-resolved spectra with the same model, it was decided to reject it.

Recently, \citet{2019ApJS..243...29A} and \citet{2021MNRAS.501.5892S} %aftab+ 2019 and graciela+ 2020
carried out spectral analysis of \xmm data in the eclipse and out-of-eclipse phases in the energy band (0.8--10.0 keV). They obtained best fits by combining blackbody and power-law components and used them to describe both eclipse and out-of-eclipse spectra. In addition, \citep{2001ApJ...547..973W} argued that the stellar wind in the system is smooth and concluded that the wind is most likely driven by X-ray heating of the illuminated surface
of the companion star as proposed by \citet{1993ApJ...403..322D}. % day&stevens 1993
Compared to previous studies, thanks to \maxi's observation strategy, a large number of complete orbits have been observed and divided into 10 orbital phase intervals.
According to the previous discussion, finally, models including either a blackbody component (\emph{bbody} in \textsc{XSPEC}) or a Comptonisation component 
(\emph{compST} in \textsc{XSPEC}) have been combined to describe the average spectrum.

Interstellar medium (ISM) absorption and local absorption components have been included by means of a partial covering fraction defined through the parameter $C$\@. The \emph{tbnew}\footnote{\href{http://pulsar.sternwarte.uni-erlangen.de/wilms/tbabs}{\url{http://pulsar.sternwarte.uni-erlangen.de/wilms/tbabs}}}
component is a recent version of the T\"{u}bingen-Boulder absorption model which updates the absorption cross sections and abundances \citep{2000ApJ...542..914W}; %Wilms et al.
the \emph{gabs} factor has been described above; \emph{po} is a simple photon power law consisting of a dimensionless photon index ($\Gamma$) and the normalisation constant (\emph{K}), the spectral photons keV$^{-1}$ cm$^{-2}$ s$^{-1}$ at 1 keV; \emph{bbody} corresponds to a blackbody model whose parameters include the temperature $kT_\mathrm{bb}$ in keV and the normalisation \emph{norm}, defined as $L_{39}/D^{2}_{10}$, where $L_{39}$ is the source luminosity
in units of 10$^{39}$ \ergs and $D^{2}_{10}$ is the distance to the source in units of 10 kpc. This model is given by equation~(\ref{ec1}).

\begin{equation}
 F(E) = (C \times \mathit{tbnew} + (1-C) \times \mathit{tbnew}) \, (\mathit{gabs} \times (\mathit{po} + \mathit{bbody} + \mathit{GL}))
  \label{ec1}
\end{equation}

where \emph{GL} represents the Gaussian functions added to account for the emission lines. Here, parameters derived from \asca data by \citet{1996PASJ...48..425E} %Ebisawa et al.
were used to describe the Fe K$\alpha$ complex.

\begin{table}[!t]\centering
  \setlength{\tabnotewidth}{0.5\columnwidth}
  \tablecols{3}
  % Stretch the space between table columns 
  \setlength{\tabcolsep}{2.8\tabcolsep}
  \caption{Best-fit model parameters for the averaged spectrum\tabnotemark{a}} \label{tab:15}
 \begin{tabular}{lrr}
    \toprule
    Component & \multicolumn{1}{c}{Parameter} & \multicolumn{1}{c}{Value} \\
    \midrule
    P.c.f.  & $C$ & 0.71 $\pm$ 0.08 \\
\emph{tbnew} & $N^{1}_{H}$ [10$^{22}$ atoms cm$^{-2}$] & 19$^{+4}_{-3}$ \\
 & $N^{2}_{H}$ [10$^{22}$ atoms cm$^{-2}$] & 2.5 $^{+0.6}_{-0.5}$ \\ \midrule
\emph{gabs} & E[keV] & 5.12 $\pm$ 0.05 \\
 & $\sigma$ [keV] & 0.02 (\emph{frozen}) \\
 & line depth & 0.69 $^{+1.2}_{-0.4}$ \\ \midrule
\emph{Power law} & Photon index $\Gamma$ & 2.16$\pm$0.15 \\
 & norm [keV$^{-1}$ s$^{-1}$ cm$^{-2}$] & 0.44 $^{+0.14}_{-0.11}$ \\ \midrule
\emph{bbody} & $kT$ [keV] & 3.46 $\pm$ 0.04 \\
 & norm  [$L_{39}/D^{2}_{10}$] & 0.0171$^{+0.0021}_{-0.0018}$\\ \midrule
Fe K$\alpha$ & Line E [keV]& 6.42 (\emph{frozen}) \\
 & $\sigma$ [keV] & 0.01 (\emph{frozen}) \\
 & EW [keV]& 0.023 $\pm$ 0.004 \\
 & norm [10$^{-4}$ ph s$^{-1}$ cm$^{-2}$]& 8.0$\pm$1.3 \\ \midrule
\ion{Fe}{xxv} & Line E [keV]& 6.69 (\emph{frozen}) \\
 & $\sigma$ [keV]& 0.01 (\emph{frozen}) \\
 & EW [keV]& 0.025 $\pm$ 0.004 \\
 & norm [10$^{-4}$ ph s$^{-1}$ cm$^{-2}$]& 8.4$\pm$1.3 \\ \midrule
\ion{Fe}{xxvi} & Line E [keV]& 6.99 (\emph{frozen}) \\
 & $\sigma$ [keV] & 0.01 (\emph{frozen}) \\
 & EW [keV]& 0.017$\pm$0.004 \\
 & norm [10$^{-4}$ ph s$^{-1}$ cm$^{-2}$]& 5.2$\pm$1.3 \\\midrule
$\chi^{2}_{r}$ & \multicolumn{2}{c}{$\chi^{2}$/(d.o.f.) = 115/106 = 1.1}  \\
    \bottomrule
    \tabnotetext{a}{Parameters for equation~(\ref{ec1}). P.c.f. is the partial covering fraction. EW represents the equivalent width of the emission line. Uncertainties are given at the 90\% ($\Delta \chi^2 = 2.71$) confidence limit and d.o.f is degrees of freedom.}
  \end{tabular}
\end{table}

Fitted parameters for the continuum model are listed in Table~\ref{tab:15} where it is also included the equivalent width (EW) of the Gaussian emission lines. Figure~\ref{figure1} shows the data, the best-fit model described by equation~(\ref{ec1}), and residuals as the difference between observed flux and model flux divided by the uncertainty of the observed flux.

Since the surface luminosity of a blackbody only depends on its temperature, it is possible to calculate the radius of the emitting region, $R_\mathrm{bb}$, by using the expression:
\begin{equation}
R_\mathrm{bb}(\mathrm{km})=3.04 \times 10^{4} \frac{D\sqrt{F_\mathrm{bb}}}{T^{2}_\mathrm{bb}},
\label{ec2}
\end{equation}

where $D$ is the distance to the source in kpc, $F_\mathrm{bb}$ is the unabsorbed flux in erg s$^{-1}$ cm$^{-2}$ in the energy range 2.0--20.0 keV, $T_\mathrm{bb}$ the temperature in keV. Taking into account the distance to the source given by \emph{GAIA (E)DR3} $d (kpc) = 6.8^{+0.6}_{-0.5}$, and an unabsorbed flux of $(1.7^{+0.5}_{-0.4}) \times 10^{-9}$ \ergscm in the \maxi energy band, from equation~(\ref{ec2}) we found a radius of the emitting surface of $R_\mathrm{bb} = 0.71^{+0.19}_{-0.16}$ km. If we assume thermal emission from the NS polar cap, this radius may be consistent with the expected size \citep[also compatible with][]{2021MNRAS.501.5892S}. %G. Sanjurjo et al., 2020

The intrinsic bolometric X-ray luminosity is a key parameter to infer the stellar wind parameters and the details of the accretion processes. It is usually estimated from the measured X-ray flux of the source by using the equation~(\ref{ec3}). Although the accretion flow is not expected to be as isotropic as the stellar wind, here, it is assumed that the system is emitting isotropically. A few cautions should be kept in mind when using this assumption: in accreting neutron stars, the bulk of the x-rays are produced in the accretion columns near the two magnetic poles; the emission profiles of these regions are not well known; the bolometric flux is measured on a small energy band and derived from phenomenological rather than physically justified models (see \citet{2017SSRv..212...59M} for a review of stellar winds from massive stars).

\begin{equation}
L_\mathrm{X}=4\, \pi\, D^{2}\, f_\mathrm{no\_abs},
\label{ec3}
\end{equation}

where $L_\mathrm{X}$ is the X-ray luminosity, $D$ is the distance to the source and $f_\mathrm{no\_abs}$ is the unabsorbed flux in the 2.0--20.0 keV energy band. We obtained $L_\mathrm{X} = (1.9^{+1.0}_{-0.8})\times10^{37}$ \ergs ($f_\mathrm{no\_abs} = (3.5^{+1.1}_{-0.9})\times 10^{-9}$ \ergscm) indicating that the accretion mode is not only due to the stellar wind ($>$ 10$^{36}$ erg s$^{-1}$) but should also be enhanced by disk accretion or gas stream accretion.

From the \emph{bbody} normalization, $L_{39}/D^{2}_{10}$, a value of $0.04^{+0.09}_{-0.08}$ is derived, which is consistent with the experimental result if we take into account the uncertainties of the parameters.

Another model has been also tested by replacing the blackbody by a comptonisation model, maintaining the rest of the components unchanged. This model is described by equation~(\ref{ec4}).

\begin{equation}
  F(E) = (C \times \mathit{tbnew} + (1-C) \times \mathit{tbnew}) \, (gabs \times (\mathit{po} + \mathit{compST} + \mathit{GL}))
  \label{ec4}
\end{equation}

\begin{figure}[hbtp]
  \centering
  \includegraphics[angle=-90.0,width=\columnwidth]{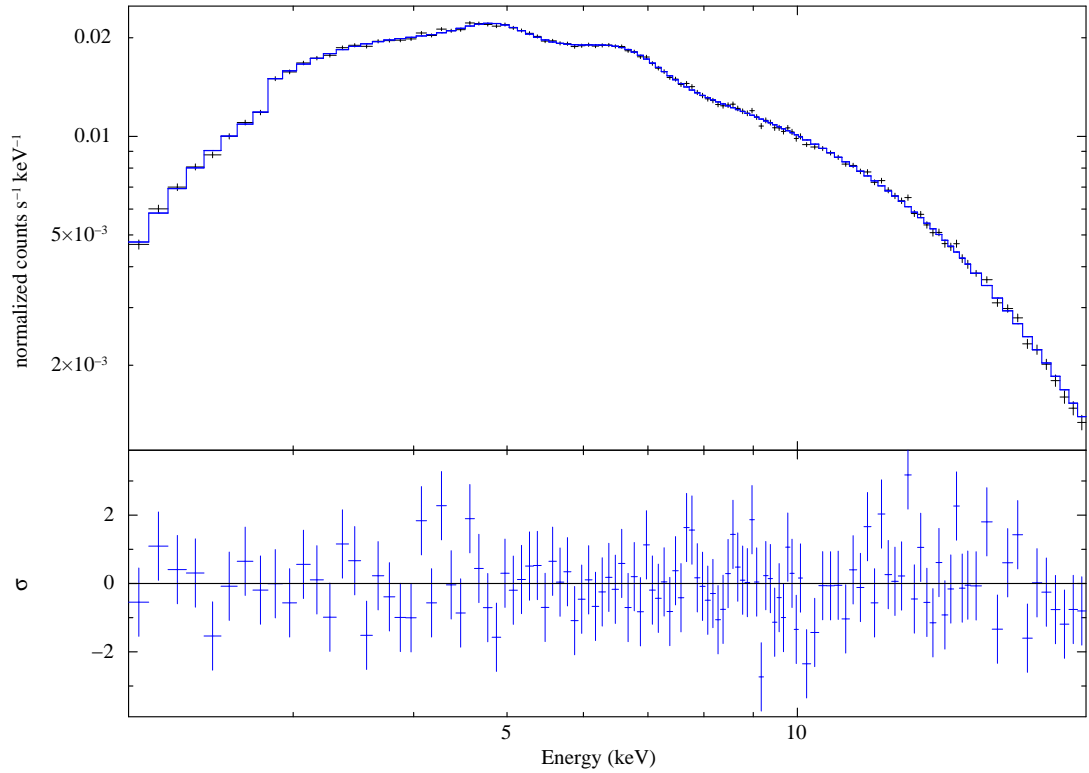}  
  \caption[]{\label{figure2} %
     Orbital phase averaged spectra of \centau
    in the 2.0--20.0 keV band.
    \emph{Top panel}: data and 
     best-fit model (equation~(\ref{ec4})).
    \emph{Bottom panel} shows the residuals between the spectrum and the model. The resulting fit parameters are reported in Table~\ref{tab:16}.   
  }
\end{figure}

Best-fit model parameters are listed in Table~\ref{tab:16} and Figure~\ref{figure2} shows the averaged spectrum. From the model, the inferred unabsorbed flux was
$(3^{+6}_{-1})\times 10^{-9}$ \ergscm in the \maxi energy band, corresponding to an X-ray luminosity of $L_\mathrm{X} = (2^{+3}_{-1})\times 10^{37}$ \ergs which agrees completely with the previous result.

\begin{table}[!t]\centering
  \setlength{\tabnotewidth}{0.5\columnwidth}
  \tablecols{3}
  % Stretch the space between table columns 
  \setlength{\tabcolsep}{2.8\tabcolsep}
  \caption{Best-fit model parameters for the averaged spectrum\tabnotemark{a}} \label{tab:16}
 \begin{tabular}{lrr}
    \toprule
    Component & \multicolumn{1}{c}{Parameter} & \multicolumn{1}{c}{Value} \\
    \midrule
    P.c.f. & $C$ & 0.80$^{+0.09}_{-0.08}$ \\
\emph{tbnew} & $N^{1}_{H}$ [10$^{22}$ atoms cm$^{-2}$] & 15$^{+4}_{-3}$ \\
 & $N^{2}_{H}$ [10$^{22}$ atoms cm$^{-2}$] & 1.9$^{+1.2}_{-0.8}$ \\ \midrule
\emph{gabs} & E[keV] & 5.12$\pm$0.05 \\
 & $\sigma$ [keV] & 0.02 (\emph{frozen}) \\
 & line depth & 0.5$^{+0.8}_{-0.3}$ \\ \midrule
\emph{Power law} & $\Gamma$ & 2.6$^{+0.8}_{-0.5}$ \\
 & norm [keV$^{-1}$ s$^{-1}$ cm$^{-2}$] & 0.3$^{+0.5}_{-0.1}$ \\ \midrule
\emph{compST} & kT [keV] & 3.935$\pm$0.009 \\
 & $\tau$ & 18.6 $^{+0.8}_{-1.0}$ \\
 & norm & 0.08 $^{+0.03}_{-0.02}$ \\ \midrule
Fe K$\alpha$ & Line E [keV]& 6.42 (\emph{frozen}) \\
 & $\sigma$ [keV] & 0.01 (\emph{frozen}) \\
 & EW [keV]& 0.035$\pm$0.004 \\
 & norm [10$^{-3}$ ph s$^{-1}$ cm$^{-2}$]& 1.05$\pm$0.11 \\ \midrule
\ion{Fe}{xxv} & Line E [keV]& 6.69 (\emph{frozen}) \\
 & $\sigma$ [keV]& 0.01 (\emph{frozen}) \\
 & EW [keV]& 0.016$\pm$0.004 \\
 & norm [10$^{-4}$ ph s$^{-1}$ cm$^{-2}$]& 4.8$\pm$1.1 \\ \midrule
\ion{Fe}{xxvi} & Line E [keV]& 6.99 (\emph{frozen}) \\
 & $\sigma$ [keV] & 0.01 (\emph{frozen}) \\
 & EW [keV]& 0.021$\pm$0.004 \\
 & norm [10$^{-4}$ ph s$^{-1}$ cm$^{-2}$]& 5.2$\pm$1.1 \\ \midrule
$\chi^{2}_{r}$ & \multicolumn{2}{c}{$\chi^{2}$/(d.o.f.) = 112/105 = 1.1}  \\
    \bottomrule
    \tabnotetext{a}{Parameters for equation~(\ref{ec4}). P.c.f. is the partial covering fraction. EW represents the equivalent width of the emission line. Uncertainties are given at the 90\% confidence limit and d.o.f is degrees of freedom.}
  \end{tabular}
\end{table}

The Comptonisation parameter $y = k\, T\, \tau^{2}/(m_{e}\, c^{2})$, where $k$ is the Boltzmann constant, $T$ is the temperature, $\tau$ is the optical depth, m$_{e}$ is the electron mass and $c$ is the light speed, determines the efficiency of the Comptonisation process \citep{1994ApJ...434..570T,2008MNRAS.389..301P}, and its value from the averaged spectrum $y = 2.7^{+0.2}_{-0.3}$ indicates an efficient process that corresponds to a moderate accretion rate.

The iron fluorescence emission line shows an interesting evolution along the orbital phase: it is centred at 6.4 keV out-of-eclipse but it is shifted to 6.7 keV at the ingress suggesting the coexistence of two iron lines at different energies \citep{2017symm.conf..159R}. Sensitive X-ray observatories such as \emph{ASCA} and \emph{XMM-Newton} have detected and resolved the Fe complex in \centau \citep[][ respectively]{1996PASJ...48..425E,2021MNRAS.501.5892S}. In contrast, \maxi is not able to resolve it and therefore the values obtained for these lines by \citet{1996PASJ...48..425E} %Ebisawa et al.
have been used to fit them. For this purpose, all their parameters except the line intensities were fixed (see Tables~\ref{tab:15} and ~\ref{tab:16}). The line flux ratio
[\ion{Fe}{xxvi}]/[\ion{Fe}{xxv}] can be used to estimate the ionisation state of the emitting plasma. The results obtained in the averaged spectrum for both models were $0.6\pm0.3$ (equation~(\ref{ec1})) and $1.1\pm0.5$ (equation~(\ref{ec4})), respectively. Assuming that this procedure is an approximation of the ionisation state of the system, these average values point to a highly ionised plasma with $\log \xi \sim (3.4-3.8)$, according to the ionisation parameter calculated by \citet[their Figure 8, upper panel]{1996PASJ...48..425E}. %Ebisawa et al.

Since the partial covering fraction modifies the continuum at low energies, two hydrogen column components were applied: one to describe the ISM towards the system, $N^{2}_{H}$, and the other to describe the ISM plus local absorption, $N^{1}_{H}$. The mean value of the Galactic column density of Hydrogen $N_{H,tot}$ in the direction
of \centau is $1.16 \times$ 10$^{22}$ cm$^{-2}$ \citep{2013MNRAS.431..394W}. %Willingale et al.
 \citet{2015ApJ...809...66V} % valencic & smith
reported an ISM absorption towards this source of $(1.6\pm0.3) \times$ 10$^{22}$ cm$^{-2}$. The column $N^{2}_{H}$ derived from both models are compatible taking into account its uncertainties.

\subsection{Orbital phase averaged, high and low states spectra with MAXI/GSC}
\label{averaged_spectrum_8}

\begin{figure}[hbtp]
  \centering
  \includegraphics[angle=-90.0,width=\columnwidth]{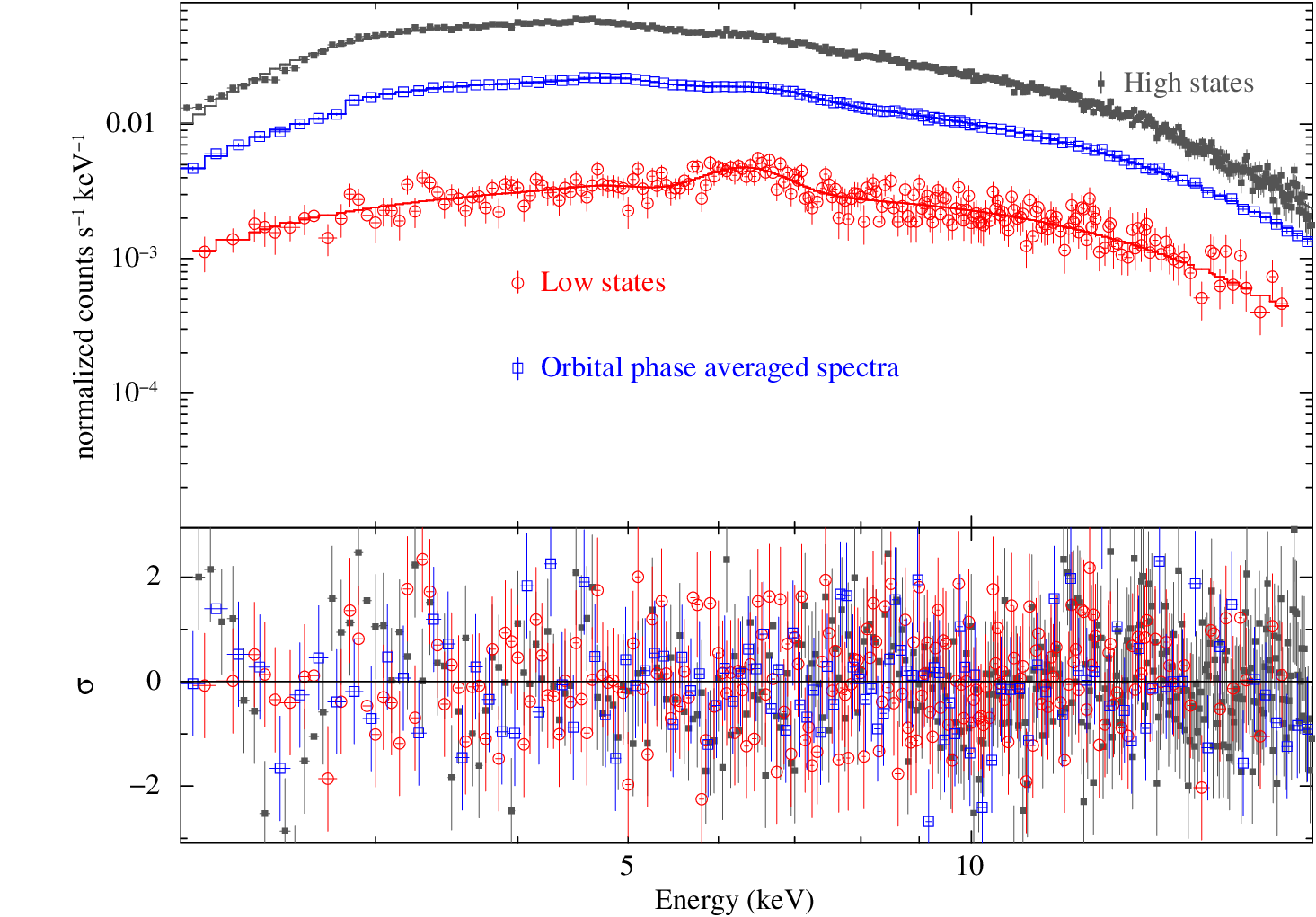}  
  \caption[]{\label{figure40} %
    Orbital phase averaged, high states and low states spectra of \centau
    in the 2.0--20.0 keV band.
    \emph{Top panel}: Data and best-fit models (defined by equation~(\ref{ec1})).
    \emph{Bottom panel} shows the residuals for the model. The resulting fit parameters are reported in Table~\ref{tab:21}.   
  }
\end{figure}

\begin{figure}[hbtp]
  \centering
  \includegraphics[angle=-90.0,width=\columnwidth]{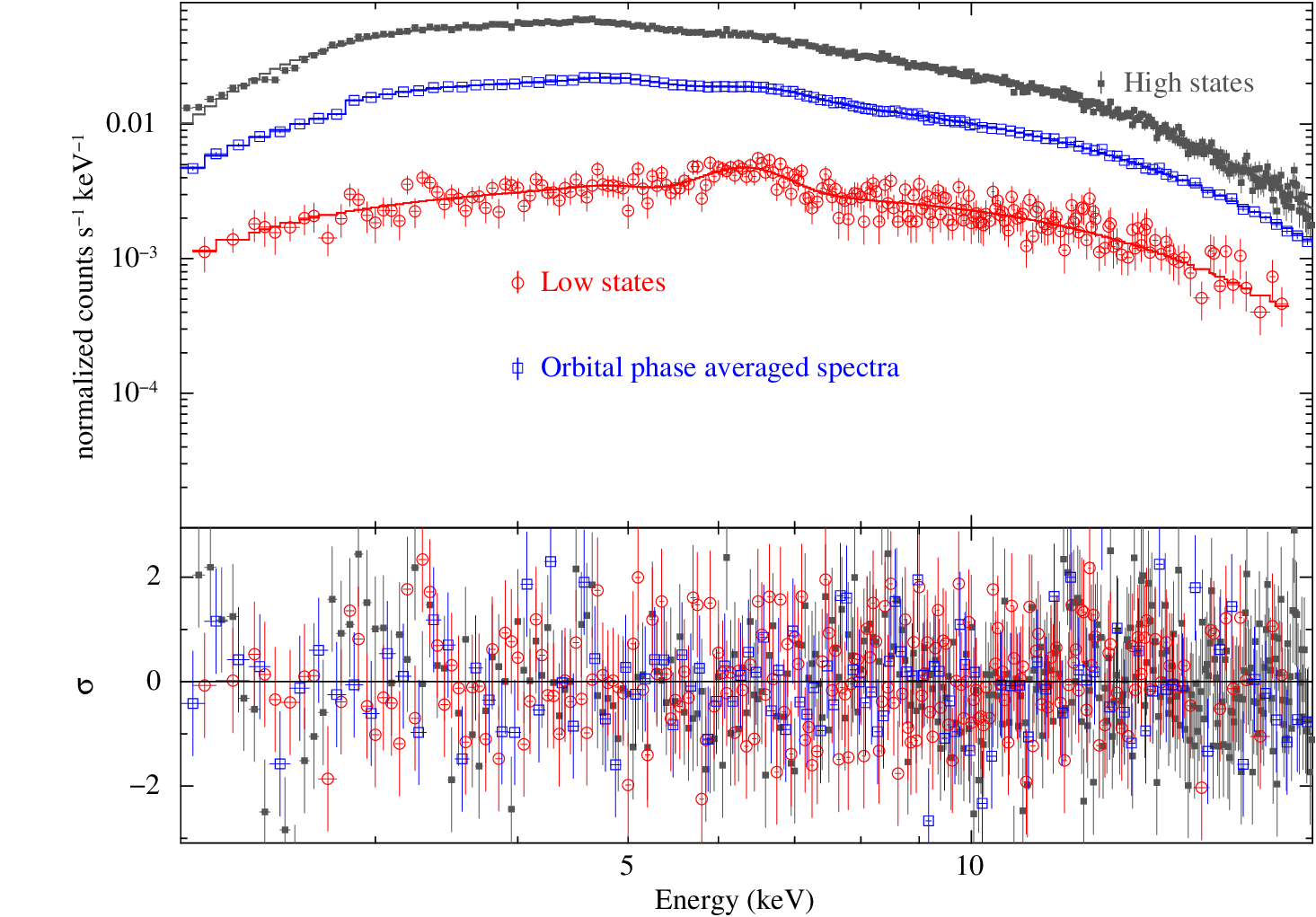}  
  \caption[]{\label{figure41} %
    Orbital phase averaged, high states and low states spectra of \centau
    in the 2.0--20.0 keV band.
    \emph{Top panel}: Data and best-fit models (defined by equation~(\ref{ec4})).
    \emph{Bottom panel} shows the residuals for the model.     The resulting fit parameters are reported in Table~\ref{tab:22}.
  }
\end{figure}

As a consequence of the large variability of the light curve, high and low states have been defined. Thus, the good time intervals (GTIs) of both states have been identified in the energy range 2.0--20.0 keV and their respective averaged spectra have been extracted. In this analysis, we fitted the high and low states spectra with  the same models as we used in the orbital phase-averaged spectra spectrum, except that only a single Gaussian has been included here. The fits are shown in Figures~\ref{figure40} and ~\ref{figure41} meanwhile the best-fit model parameters are listed in Tables~\ref{tab:21} (equation~(\ref{ec1})) and \ref{tab:22} (equation~(\ref{ec4})).

To determine the periods of high and low activity we established a number of counts greater than $\sim$0.5 photons s$^{-1}$ cm$^{-2}$ and less than $\sim$0.3 photons s$^{-1}$ cm$^{-2}$, respectively. Then, more than 100 bins (one bin $\sim 0.1$ MJD) were accumulated to calculate the low-activity intervals and grouped 10 bins by 10 bins for the high-activity events.

The unabsorbed fluxes of the orbital phase-averaged, high and low activity spectra in units of \ergscm are $(1.72\pm 0.03)\times10^{-9}$, $(4.1^{+1.3}_{-1.0})\times10^{-9}$ and $(5.4^{+1.5}_{-1.2}) \times 10^{-10}$, respectively. For such fluxes, the radius of the blackbody emitting area is found $R_\mathrm{bb} = 0.71^{+0.07}_{-0.06}$ km (averaged spectra), $R_\mathrm{bb} = 1.6\pm0.5$ km (high states) and $R_\mathrm{bb} = 0.36^{+0.12}_{-0.10}$ km (low states). All values are on the order of 1 to 2 km which is compatible with a hotspot on the NS surface \citep{2021MNRAS.501.5892S}.

Using the definition of the \emph{bbody} normalization, $L_{39}/D^{2}_{10}$, we have derived its value in the three spectra: $0.04^{+0.08}_{-0.07}$ (averaged spectra), $0.1\pm0.3$ (high states) and $0.009^{+0.019}_{-0.016}$ (low states). These values are consistent with the experimental ones, taking the uncertainties into account (see Table~\ref{tab:21}).

The Comptonisation parameter $y$ (see Table~\ref{tab:23}) shows an efficient process that corresponds to a moderate accretion rate if we take into account the uncertainties.

During the high state, when $L_X$ is close to 10$^{38}$ erg s$^{-1}$, the Fe K$\alpha$ is not present in the spectrum. Moreover, the EW of the Fe K$\alpha$ is three times higher during the low states than that in the averaged spectra. These facts suggest that the strong x-ray radiation may mask the possible presence of iron emission lines.

The unabsorbed flux is found to vary with similar trend for both models. In fact, the X-ray luminosity is the same, taking the uncertainties into account (see Table~\ref{tab:23}, column 2 corresponds to equation~(\ref{ec1}) and column 3 corresponds to equation~(\ref{ec4})). On the one hand, it indicates (for averaged spectra and high states) that the accretion mode is not only due to the stellar wind ($>$ 10$^{36}$ erg s$^{-1}$) but should also be enhanced by disk accretion or gas stream accretion. On the other hand, the luminosity indicates that the accretion mode in low states spectra is due to stellar wind. Therefore, the difference in X-ray luminosity between the high and low states can be attributed to a decrease in the accretion rate rather than an overall rise in absorption.

Tables~\ref{tab:28} and ~\ref{tab:29} show the unabsorbed flux and the luminosity of each model component (equations~(\ref{ec1}) and (\ref{ec4}), respectively) as well as the total unabsorbed fluxes and luminosities whose values agree with those given in Table~\ref{tab:23}.

{\scriptsize
\begin{table}[!t]\centering
{\scriptsize
  \setlength{\tabnotewidth}{0.5\columnwidth}
  \tablecols{3}
  % Stretch the space between table columns 
  \setlength{\tabcolsep}{2.8\tabcolsep}
  \caption{Best-fit model parameters for the averaged, high states and low states spectra\tabnotemark{a}} \label{tab:21}
 \begin{tabular}{lrr}
    \toprule
    Component & \multicolumn{1}{c}{Parameter} & \multicolumn{1}{c}{Value} \\
    \midrule
    P.c.f.  & $C$ & 0.809 $\pm$ 0.003 \\
\emph{tbnew} & $N^{1}_{H}$ [10$^{22}$ atoms cm$^{-2}$] & 18.8 $\pm$ 0.6 \\
 & $N^{2}_{H}$ [10$^{22}$ atoms cm$^{-2}$] & 2.57 $\pm$ 0.07 \\ \midrule
\emph{gabs} & E[keV] & 5.12 (\emph{frozen}) \\
 & $\sigma$ [keV] & 0.02 (\emph{frozen}) \\
 & line depth & 0.69 (\emph{frozen}) \\ \midrule
\emph{Power law} & $\Gamma$ & 2.184$\pm$0.009 \\
 & norm [keV$^{-1}$ s$^{-1}$ cm$^{-2}$] & 0.405$\pm$ 0.006 \\ \midrule
\emph{bbody} & $kT$ [keV] & 3.4628$\pm$0.0023 \\
 & norm  [$L_{39}/D^{2}_{10}$] & 0.01527$\pm$0.00014\\ \midrule
Fe K$\alpha$ & Line E [keV]& 6.67$\pm$0.05 \\
 & $\sigma$ [keV] & 0.28$^{+0.07}_{-0.08}$ \\
 & EW [keV]& 0.070$\pm$0.004 \\
 & norm [10$^{-3}$ ph s$^{-1}$ cm$^{-2}$]& 1.98 $\pm$ 0.12 \\\midrule
$\chi^{2}_{r}$ & \multicolumn{2}{c}{$\chi^{2}$/(d.o.f.) = 115/108 = 1.1}  \\\midrule\midrule
P.c.f.  & $C$ & 0.800 $\pm$ 0.003 \\
\emph{tbnew} & $N^{1}_{H}$ [10$^{22}$ atoms cm$^{-2}$] & 63$^{+17}_{-11}$ \\
 & $N^{2}_{H}$ [10$^{22}$ atoms cm$^{-2}$] & 6.9$\pm$0.5  \\ \midrule
\emph{gabs} & E[keV] & 5.12 (\emph{frozen}) \\
 & $\sigma$ [keV] & 0.02 (\emph{frozen}) \\
 & line depth & 0.69 (\emph{frozen}) \\ \midrule
\emph{Power law} & $\Gamma$ & 2.65$^{+0.22}_{-0.21}$ \\
 & norm [keV$^{-1}$ s$^{-1}$ cm$^{-2}$] & 3.8$^{+1.2}_{-0.9}$ \\ \midrule
\emph{bbody} & $kT$ [keV] & 2.87$^{0.09}_{-0.18}$ \\
 & norm  [$L_{39}/D^{2}_{10}$] & 0.033$\pm$0.004\\\midrule
$\chi^{2}_{r}$ & \multicolumn{2}{c}{$\chi^{2}$/(d.o.f.) = 452/351 = 1.3}  \\\midrule\midrule
P.c.f.  & $C$ & 0.80$\pm$0.03 \\
\emph{tbnew} & $N^{1}_{H}$ [10$^{22}$ atoms cm$^{-2}$] & 18$^{+7}_{-5}$ \\
 & $N^{2}_{H}$ [10$^{22}$ atoms cm$^{-2}$] & 1.4$^{+0.9}_{-0.7}$ \\ \midrule
\emph{gabs} & E[keV] & 5.12 (\emph{frozen}) \\
 & $\sigma$ [keV] & 0.02 (\emph{frozen}) \\
 & line depth & 0.69 (\emph{frozen}) \\ \hline
\emph{Power law} & $\Gamma$ & 2.55$\pm$0.18 \\
 & norm [keV$^{-1}$ s$^{-1}$ cm$^{-2}$] & 0.071$^{+0.020}_{-0.016}$ \\ \midrule
\emph{bbody} & $kT$ [keV] & 3.67$^{+0.18}_{-0.17}$ \\
 & norm  [$L_{39}/D^{2}_{10}$] & 0.0050$^{+0.0004}_{-0.0003}$\\ \midrule
Fe K$\alpha$ & Line E [keV]& 6.31$\pm$0.10 \\
 & $\sigma$ [keV] & 0.22$^{+0.19}_{-0.22}$ \\
 & EW [keV]& 0.22 $\pm$ 0.04 \\
 & norm [10$^{-3}$ ph s$^{-1}$ cm$^{-2}$]& 1.6$\pm$0.3 \\\midrule
$\chi^{2}_{r}$ & \multicolumn{2}{c}{$\chi^{2}$/(d.o.f.) = 206/220 = 0.9}  \\
    \bottomrule
    \tabnotetext{a}{{\scriptsize Parameters for equation~(\ref{ec1}). P.c.f. is the partial covering fraction. EW represents the equivalent width of the emission line. Uncertainties are given at the 90\% confidence limit and d.o.f is degrees of freedom.}}
  \end{tabular}
} 
\end{table}
}

{\scriptsize
\begin{table}[!t]\centering
{\scriptsize
  \setlength{\tabnotewidth}{0.5\columnwidth}
  \tablecols{3}
  % Stretch the space between table columns 
  \setlength{\tabcolsep}{2.8\tabcolsep}
  \caption{Best-fit model parameters for the averaged, high states and low states spectra\tabnotemark{a}} \label{tab:22}
 \begin{tabular}{lrr}
    \toprule
    Component & \multicolumn{1}{c}{Parameter} & \multicolumn{1}{c}{Value} \\
    \midrule
    P.c.f. & $C$ & 0.813$\pm$0.004 \\
\emph{tbnew} & $N^{1}_{H}$ [10$^{22}$ atoms cm$^{-2}$] & 15.9$\pm$0.5 \\
 & $N^{2}_{H}$ [10$^{22}$ atoms cm$^{-2}$] & 2.22$\pm$0.08 \\ \midrule
\emph{gabs} & E[keV] & 5.12 (\emph{frozen}) \\
 & $\sigma$ [keV] & 0.02 (\emph{frozen}) \\
 & line depth & 0.5 (\emph{frozen}) \\ \midrule
\emph{Power law} & $\Gamma$ & 2.612$\pm$0.023 \\
 & norm [keV$^{-1}$ s$^{-1}$ cm$^{-2}$] & 0.323$\pm$0.011 \\ \midrule
\emph{compST} & kT [keV] & 3.90$\pm$0.04 \\
 & $\tau$ & 19.26$^{+0.24}_{-0.23}$ \\
 & norm & 0.0709$\pm$0.0013 \\ \midrule
Fe K$\alpha$ & Line E [keV]& 6.65 $\pm$0.04 \\
 & $\sigma$ [keV] & 0.28$\pm$0.07\\
 & EW [keV]& 0.074$\pm$0.004 \\
 & norm [10$^{-3}$ ph s$^{-1}$ cm$^{-2}$]& 2.04$\pm$0.12 \\ \midrule
$\chi^{2}_{r}$ & \multicolumn{2}{c}{$\chi^{2}$/(d.o.f.) = 111/107 = 1.0}  \\\midrule\midrule
     P.c.f. & $C$ & 0.840$^{+0.012}_{-0.011}$ \\
\emph{tbnew} & $N^{1}_{H}$ [10$^{22}$ atoms cm$^{-2}$] & 66$^{+4}_{-3}$ \\
 & $N^{2}_{H}$ [10$^{22}$ atoms cm$^{-2}$] & 6.87$^{+0.12}_{-0.11}$ \\ \midrule
\emph{gabs} & E[keV] & 5.12 (\emph{frozen}) \\
 & $\sigma$ [keV] & 0.02 (\emph{frozen}) \\
 & line depth & 0.5 (\emph{frozen}) \\ \midrule
\emph{Power law} & $\Gamma$ & 2.603$\pm$0.012 \\
 & norm [keV$^{-1}$ s$^{-1}$ cm$^{-2}$] & 3.42$\pm$0.07 \\ \midrule
\emph{compST} & kT [keV] & 2.85$\pm$0.05 \\
 & $\tau$ & 48$^{+8}_{-6}$ \\
 & norm & 0.023$^{+0.006}_{-0.005}$ \\ \midrule
$\chi^{2}_{r}$ & \multicolumn{2}{c}{$\chi^{2}$/(d.o.f.) = 452/350 = 1.3}  \\\midrule\midrule
     P.c.f. & $C$ & 0.80$^{+0.04}_{-0.03}$ \\
\emph{tbnew} & $N^{1}_{H}$ [10$^{22}$ atoms cm$^{-2}$] & 18$^{+7}_{-5}$ \\
 & $N^{2}_{H}$ [10$^{22}$ atoms cm$^{-2}$] & 1.1$^{+0.9}_{-0.7}$ \\ \midrule
\emph{gabs} & E[keV] & 5.12 (\emph{frozen}) \\
 & $\sigma$ [keV] & 0.02 (\emph{frozen}) \\
 & line depth & 0.5 (\emph{frozen}) \\ \midrule
\emph{Power law} & $\Gamma$ & 2.37$\pm$0.17 \\
 & norm [keV$^{-1}$ s$^{-1}$ cm$^{-2}$] & 0.055$^{+0.015}_{-0.012}$ \\ \midrule
\emph{compST} & kT [keV] & 3.7$\pm$0.3 \\
 & $\tau$ & 43$^{+22}_{-9}$ \\
 & norm & 0.0026$^{+0.0014}_{-0.0013}$ \\ \midrule
Fe K$\alpha$ & Line E [keV]& 6.31$\pm$0.10 \\
 & $\sigma$ [keV] & 0.22$^{+0.17}_{-0.22}$ \\
 & EW [keV]& 0.22$\pm$0.03 \\
 & norm [10$^{-3}$ ph s$^{-1}$ cm$^{-2}$]& 1.61$\pm$0.23 \\ \midrule
$\chi^{2}_{r}$ & \multicolumn{2}{c}{$\chi^{2}$/(d.o.f.) = 206/219 = 0.9}  \\
    \bottomrule
    \tabnotetext{a}{{\scriptsize Parameters for equation~(\ref{ec4}). P.c.f. is the partial covering fraction. EW represents the equivalent width of the emission line. Uncertainties are given at the 90\% confidence limit and d.o.f is degrees of freedom.}}
  \end{tabular}
} 
\end{table}
}

\begin{table}%[scale=2.0,width=2\columnwidth]
 \caption{X-ray luminosity (10$^{37}$ erg s$^{-1}$) and Comptonisation parameter}
    \centering
    \begin{tabular}{lccc} \toprule\toprule
      Orbital phase & L$_{x}$ (equation~(\ref{ec1}))& L$_{x}$ (equation~(\ref{ec4})) & $y = k\, T\, \tau^{2}/(m_{e}\, c^{2})$ \\ \midrule
Averaged & $1.9^{+0.4}_{-0.3}$ & $1.9^{+0.4}_{-0.3}$ & $2.83\pm0.10$ \\
High states & $6\pm3$ & $6.3^{+1.3}_{-1.0}$ & $13^{+5}_{-3}$  \\
Low states & $0.40^{+0.18}_{-0.15}$ & $0.40^{+0.18}_{-0.14}$ & $13^{+15}_{-7}$ \\
     \bottomrule
    \end{tabular}    
    \label{tab:23}

\end{table} 

\begin{table}[h!tb]
 \caption{Unabsorbed flux (10$^{-9}$ erg s$^{-1}$ cm$^{-2}$) and luminosity (10$^{37}$ erg s$^{-1}$) of the model components for the averaged, high states and low states spectra}
    \centering
    \begin{tabular}{lll} \toprule\toprule
       Component & Unabs. flux (equation~(\ref{ec1}))& L$_{x}$  (equation~(\ref{ec1})) \\ \midrule   
\emph{Power law} & 1.72$\pm$0.03 & 0.95$^{+0.19}_{-0.15}$ \\ \midrule
\emph{bbody} & 1.72$\pm$0.03 & 0.95$^{+0.19}_{-0.15}$\\ \midrule
Fe K$\alpha$ & 0.0343$\pm$0.0005 & 0.019$^{+0.004}_{-0.003}$\\
\midrule
Total & 3.47$\pm$0.06 & 1.9$^{+0.4}_{-0.3}$\\
\midrule\midrule
\emph{Power law} & 7.3$^{+2.3}_{-1.7}$ & 4.1$\pm$2.0 \\ \midrule
\emph{bbody} & 4.1$^{+1.3}_{-1.0}$ & 2.2$^{+1.1}_{-0.9}$\\
\midrule
Total & 11$^{+4}_{-3}$ & 6$\pm$3\\
\midrule\midrule
\emph{Power law} & 0.16$\pm$0.04 & 0.09$^{+0.04}_{-0.03}$\\ \midrule
\emph{bbody} & 0.54$^{+0.15}_{-0.12}$ & 0.30$^{+0.14}_{-0.11}$\\ \midrule
Fe K$\alpha$ & 0.026$^{+0.007}_{-0.006}$ & 0.014$^{+0.007}_{-0.005}$ \\
\midrule
Total & 0.73$^{+0.20}_{-0.17}$ & 0.40$^{+0.19}_{-0.15}$\\
\bottomrule
    \end{tabular} 
    \label{tab:28}
\end{table}

\begin{table}[]
 \caption{Unabsorbed flux (10$^{-9}$ erg s$^{-1}$ cm$^{-2}$) and luminosity (10$^{37}$ erg s$^{-1}$) of the model components for the averaged, high states and low states spectra}
    \centering
    \begin{tabular}{lll} \toprule\toprule
        Component & Unabs. flux (equation~(\ref{ec4})) & L$_{x}$ (equation~(\ref{ec4}))   \\ \midrule
\emph{Power law} &0.676$\pm$0.023 &0.37$^{+0.08}_{-0.07}$\\ \midrule
\emph{compST} &2.67$\pm$0.09 & 1.5$\pm$0.3\\ \midrule
Fe K$\alpha$ &0.0353$\pm$0.0012 &0.020$^{+0.004}_{-0.003}$\\
\midrule
Total & 3.38$\pm$0.11 & 1.9$\pm$0.4\\
\midrule\midrule
\emph{Power law} &7.49$\pm$0.15 &4.2$^{+0.8}_{-0.7}$\\ \midrule
\emph{compST} &3.94$\pm$0.08 &2.2$\pm$0.4\\
\midrule
Total & 11.43$\pm$0.23 & 6.4$^{+1.2}_{-1.1}$\\
\midrule\midrule
\emph{Power law} &0.17$^{+0.05}_{-0.04}$ &0.09$^{+0.04}_{-0.03}$\\ \midrule
\emph{compST} &0.52$^{+0.14}_{-0.11}$ &0.29$^{+0.13}_{-0.10}$\\ \midrule
Fe K$\alpha$ &0.026$^{+0.007}_{-0.006}$ &0.014$^{+0.007}_{-0.005}$\\ 
\midrule
Total & 0.72$^{+0.20}_{-0.17}$ & 0.39$^{+0.18}_{-0.14}$\\
\bottomrule
    \end{tabular}
    \label{tab:29}
\end{table}

\subsection{Orbital phase-resolved spectra with MAXI/GSC}
\label{averaged_spectrum_2}

Previous studies following this direction of analysis have been performed in one or two orbits \citep{1992ApJ...396..147N,2008ApJ...675.1487S} % nagase+, suchy+
or in shorter orbital phase \citep{1996PASJ...48..425E,1996ApJ...457..397A,2001ApJ...547..973W,2019ApJS..243...29A,
2021MNRAS.501.5892S}. % ebisawa+, audley+, wojdowski+, aftab+, graciela+
We have obtained orbital phase-resolved spectra of the HMXB pulsar \centau, accumulating the 60 s duration scans into ten orbital phase bins covering entirely its orbit \citep{2015A&A...580A.140R}. % rodes-roca+

\begin{figure}[hbtp]
  \centering
  \includegraphics[angle=-90.0,width=\columnwidth]{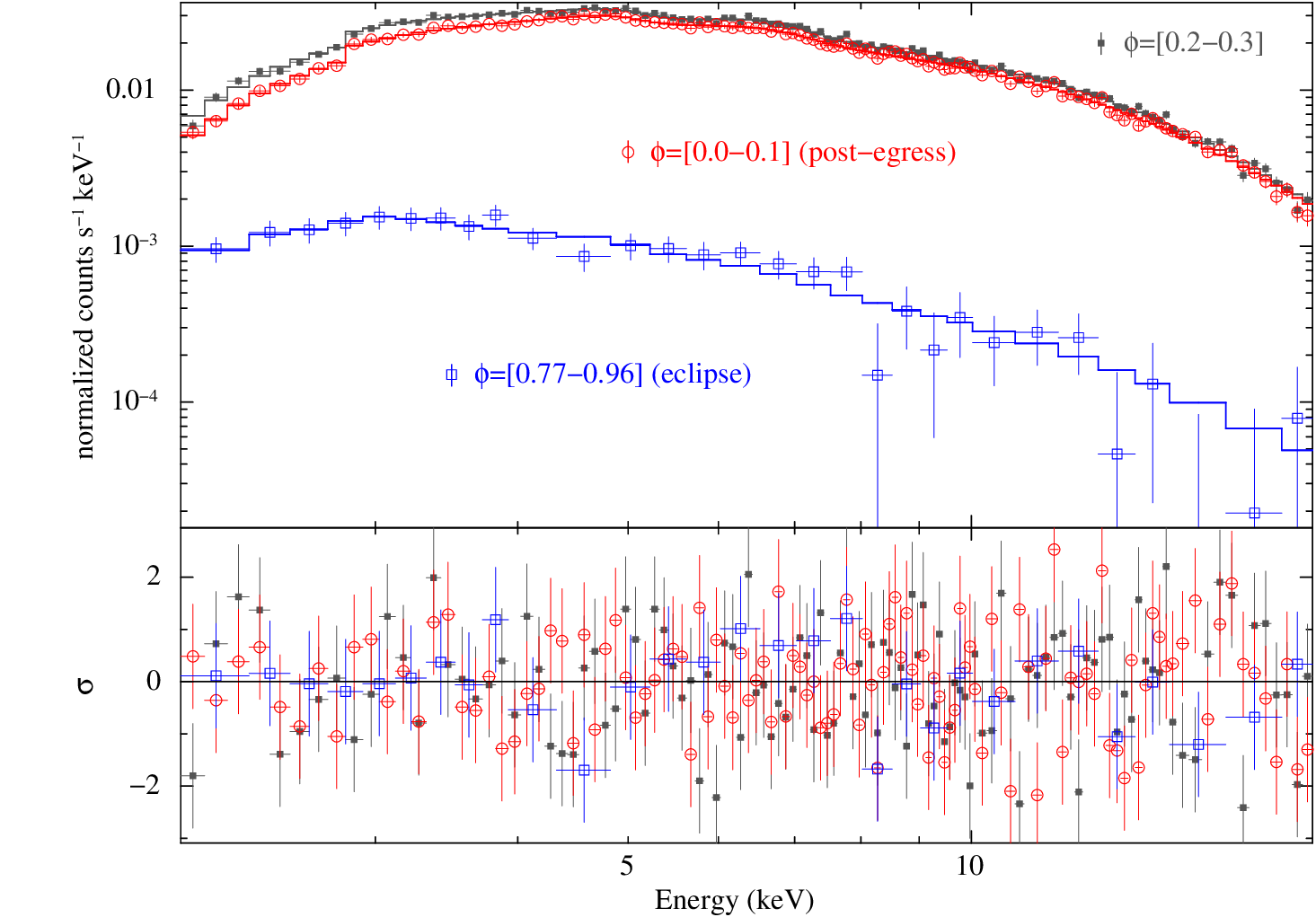} 
  \caption[]{\label{figure3} %
    Orbital phase-resolved spectra of \centau
    in the 2.0--20.0 keV band.
    \emph{Top panel}: Selected spectra and best-fit models (defined by equation~(\ref{ec1})).
    \emph{Bottom panel} shows the residuals for the model.   
  }
\end{figure}

\begin{figure}[hbtp]
  \centering
  \includegraphics[angle=-90.0,width=\columnwidth]{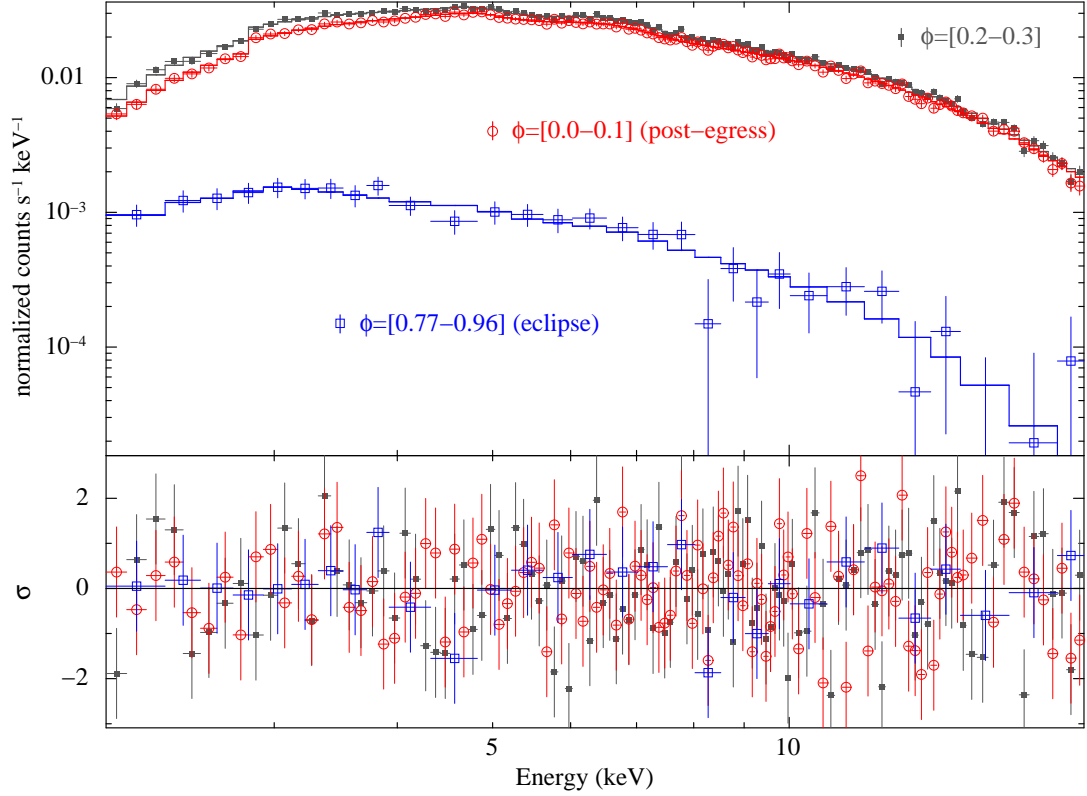}
  \caption[]{\label{figure4} %
    Orbital phase-resolved spectra of \centau
    in the 2.0--20.0 keV band.
    \emph{Top panel}: Selected spectra and best-fit models (defined by equation~(\ref{ec4})).
    \emph{Bottom panel} shows the residuals for the model.   
  }
\end{figure}

Based on the results from \S~\ref{averaged_spectrum}, we fitted the orbital phase-resolved spectra with the same two models as we used in the orbital phase-averaged spectrum. Both models gave acceptable fits to observational data and the results are shown in Figures~\ref{figure3} and ~\ref{figure4}
for selected orbital phase-resolved spectra.

\begin{figure}[hbtp]
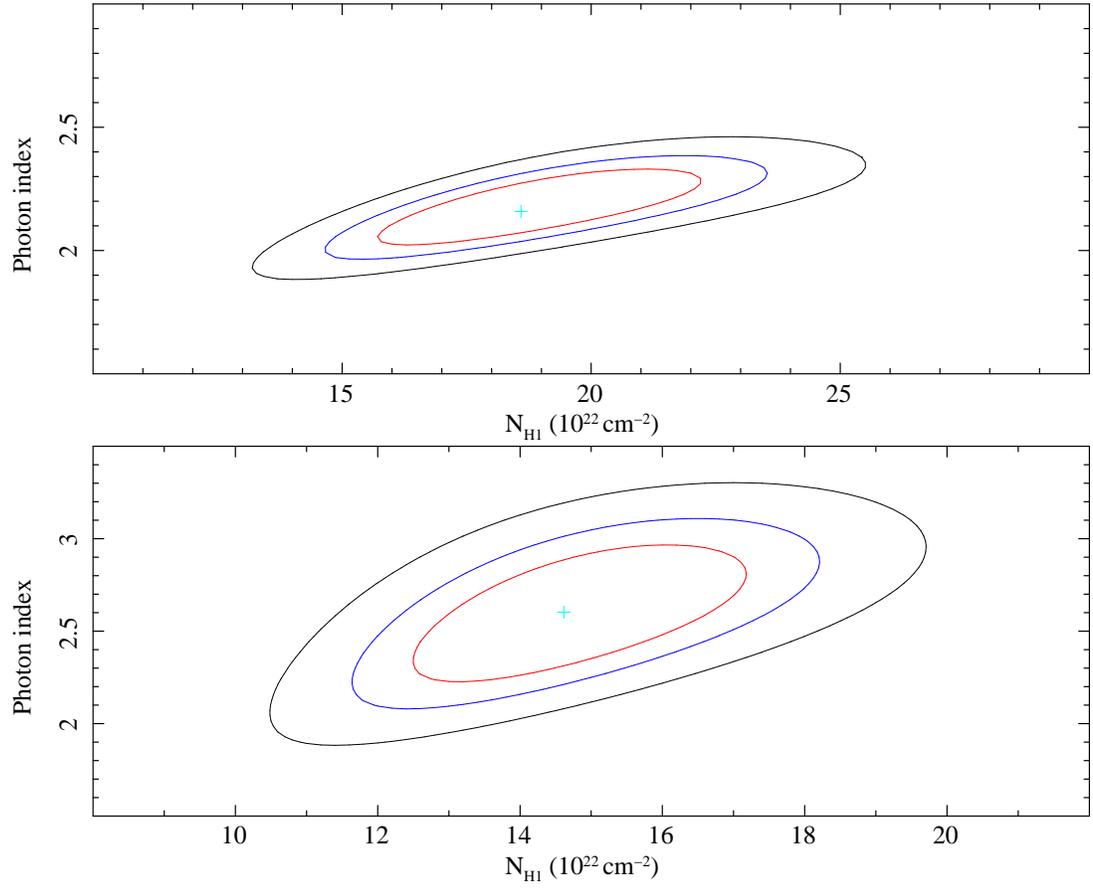

  \centering
  \includegraphics[angle=-90.0,width=\columnwidth]{elipses_bbody_espectro_medio_v8.ps}
  \includegraphics[angle=-90.0,width=\columnwidth]{elipses_compst_espectro_medio_v8.ps} 
  \caption[]{\label{figure34} % 
    The ellipses are $\chi^2$-contours for two parameters ($N_H$ and $\Gamma$). The contours are 68\%, 90\% and 99\% confidence levels for two interesting parameters. \emph{Top panel}: averaged spectrum fitted by equation~(\ref{ec1}). \emph{Bottom panel}: averaged spectrum fitted by equation~(\ref{ec4}). 
  }
\end{figure}

\begin{figure}[hbtp]
  \centering
  \includegraphics[angle=-90.0,width=\columnwidth]{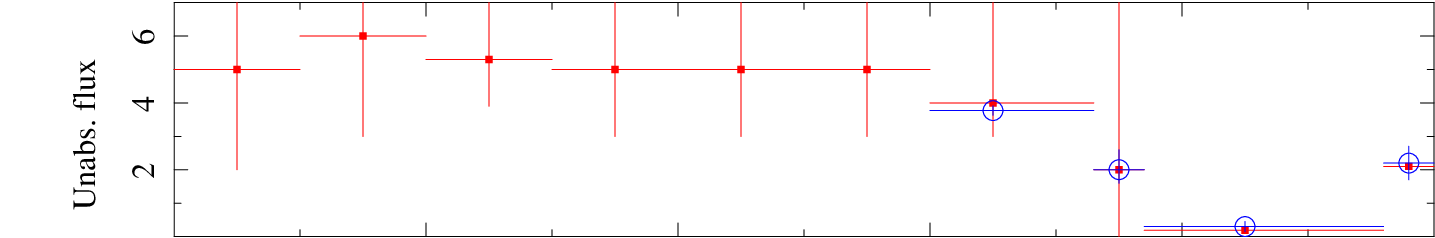} 
  \includegraphics[angle=-90.0,width=\columnwidth]{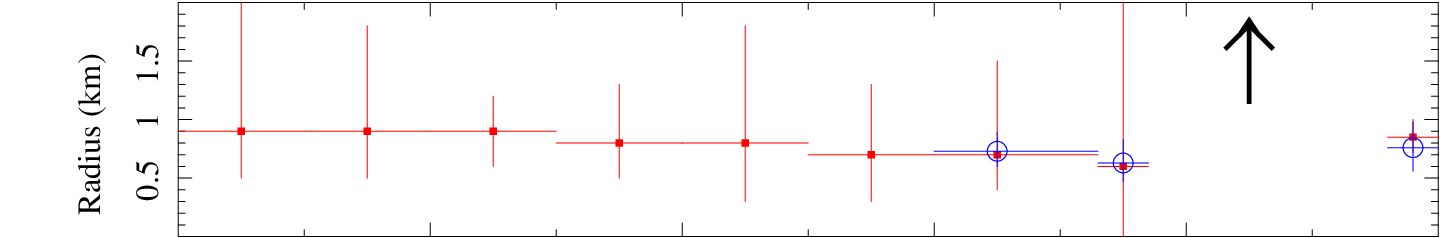} 
  \includegraphics[angle=-90.0,width=\columnwidth]{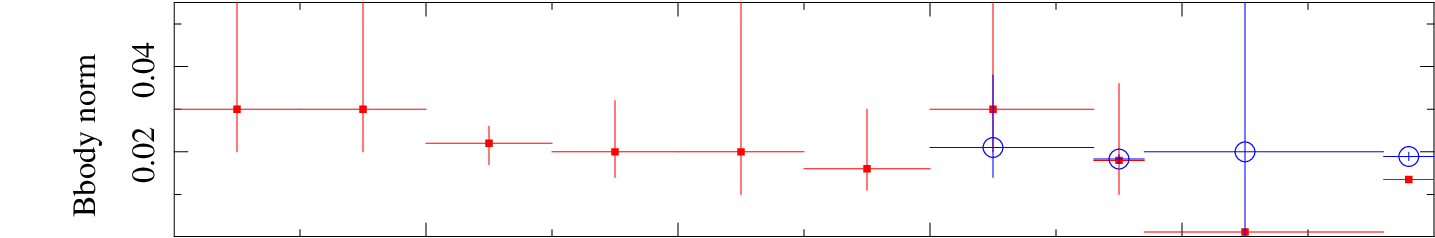}
  \includegraphics[angle=-90.0,width=\columnwidth]{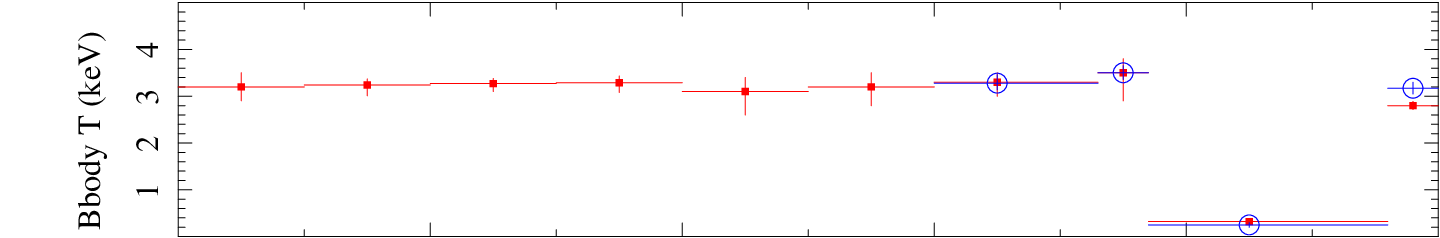} 
  \includegraphics[angle=-90.0,width=\columnwidth]{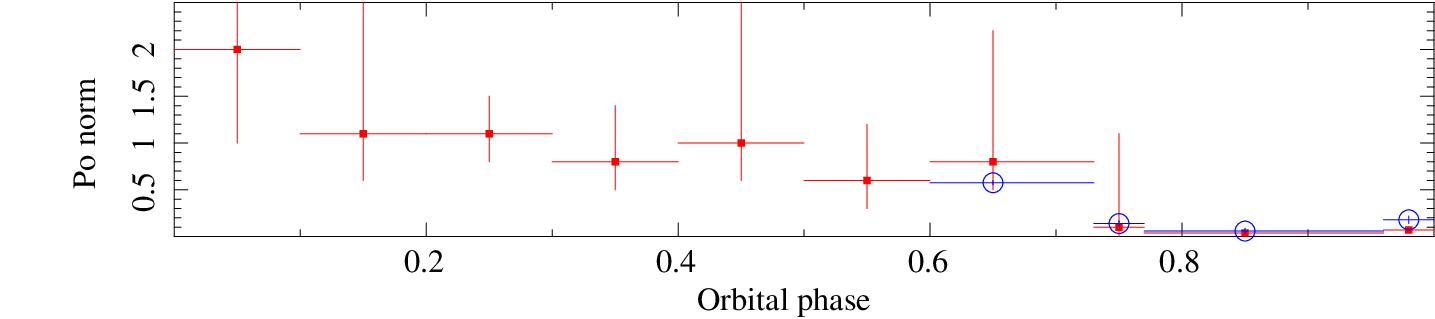} 
 \caption[]{\label{figure5} %
    Evolution of some parameters of the model described by equation~(\ref{ec1}). Unabsorbed flux is in units of 10$^{-9}$ erg cm$^{-2}$ s$^{-1}$. Bbody norm is in units of $L_{39}/D^{2}_{10}$. Power-law normalisation is expressed in units of keV$^{-1}$ s$^{-1}$ cm$^{-2}$. Filled red squares: initial values. Open blue circles: values obtained by constraining the range of values of the photon index (pre-ingress, ingress, eclipse and egress). 
}
\end{figure}

The degeneracy between the properties of the accretion rate and the physical parameters of the NS in all available models produces a certain degree of degeneracy between fit parameters. The behaviour of different parameters of the model described by the equation~(\ref{ec4}) towards eclipse suggests degeneracies between them. To better explore the photon index variation, we constrained its value to the interval obtained between orbital phase 0.0 to 0.6, i.e. $1.90 < \Gamma < 2.85$, and fitted the spectra of pre-ingress, ingress, eclipse and egress orbital phases again. The errors were obtained with the \emph{error} task, provided by XSPEC, and with propagation of uncertainties. The new values obtained by this procedure differed only slightly from the previous values within errors. However, the photon index was not well constrained and exhibited relatively large errors. Our results are marked with open blue circles in Figures~\ref{figure5}-~\ref{figure11} and ~\ref{figure12}, where filled red squares represent the initial parameter values. The possible degeneracy between the spectral photon index $\Gamma$ and absorption spectral parameters is known. Therefore, we produced $\chi^2$ contour plots of $\Gamma$ and $N_{H1}$ for both models (see Figure~\ref{figure34}). It is clear from this figure that the value of $N_{H1}$ is moderately correlated to that of $\Gamma$ as expected. However, the narrow energy band of \maxi as well as the observational mode makes it extremely difficult to remove this degeneracy through spectral fitting. In the following, we discuss the orbital phase spectral variation of the model parameters described by equations~(\ref{ec1}) and ~(\ref{ec4})
.

In Figures ~\ref{figure5}-~\ref{figure11} and \ref{figure12}-~\ref{figure31}, we show the evolution of the relevant parameters of both models throughout orbital phase. The temperature of the blackbody is almost constant, decreasing drastically in the eclipse and increasing in the eclipse-egress. That implies that the emission zone of this component should be large.
The normalisation of the blackbody component shows a smooth decrease, which could be compatible with a constant value, reaching the minimum value at eclipse (Figure~\ref{figure5}). The radius of the emission zone is of the order of 1 to 3 km whereas in eclipse it is about $9\pm3$ km (Figure~\ref{figure5}).
 This fact could suggest soft X-ray reflection from the inner accretion disk region and agree with that reported by \citet{2021MNRAS.501.5892S}. % graciela+

When the neutron star is embedded into the stellar wind of the donor, the hydrogen column density shows a modulation along the orbit. Therefore, besides the interstellar medium absorption component (consistent with a constant value), we also allowed for the presence of a local absorber, modulated by a partial covering fraction that acts as a proxy for some features of the stellar wind of the donor star or the surroundings of the compact object.

For the model defined by equation~(\ref{ec1}), the normalisation of the power law decreases smoothly during out-of-eclipse before reaching a minimum during pre-eclipse, eclipse and eclipse-egress (Figure~\ref{figure5}). The power-law photon index is rather stable during all orbital phases ($\Gamma \sim 2.1$), but it drops to 1.3 in the eclipse-egress ([0.96-1.0]) (Figure~\ref{figure12}, top panel).

For the model defined by equation~(\ref{ec4}), the normalisation of the power law (Figure~\ref{figure6}) shows a similar pattern to photon index (Figure~\ref{figure12}, fourth panel), showing a flat behaviour out-of-eclipse, increasing as the ingress takes place and decreasing at egress. Apparently, the evolution along the orbit of the power-law parameters seems to be different. However, if the range of photon index is constrained, the values obtained with the new fits become consistent with each other taking into account the uncertainties. This would suggest that there is a certain degree of degeneracy between model parameters. The orbital variation of temperature and Comptonisation normalisation have two local peaks and the lowest value occurs at the eclipse.

The orbital evolution of unabsorbed flux for both models show a very similar trend. In fact, the corresponding X-ray luminosities are the same taken into account uncertainties (see in Table~\ref{tab:17}, column 2 corresponds to equation~(\ref{ec1}) and column 3 corresponds to equation~(\ref{ec4})). The unabsorbed flux is consistent with a constant value except for eclipse and eclipse-egress. Although for an almost circular orbit and disc accretion no orbital modulation in the amount of material to accrete is expected, variations of the flux can be due to local absorption produced by an emerging accretion stream (eclipse-egress), probably corotating with the compact object, and/or a decrease in the accretion rate, most likely associated to instabilities at the inner edge of the disc interacting with the neutron star magnetosphere.

The X-ray continuum is modified by a partial covering fraction where the column $N^{2}_{H}$ corresponds to the ISM towards the source and $N^{1}_{H}$ represents the ISM plus the circumstellar environment absorption (Figure~\ref{figure12}). We have depicted the variation of the covering fraction measured from our models in Figure~\ref{figure11}. It is seen that the covering fraction tends for most of the orbit to have values lower than 0.8, $0.4 < C < 0.9$, which means that there are large inhomogeneities in the stellar wind of the giant star. Although both models seem to follow the same trend in the covering fraction, the orbital variation of $C$, given by equation~(\ref{ec4}), shows a certain modulation while the error bars obtained in equation~(\ref{ec1}) prevent us to confirm this orbital modulation. Nevertheless, taking into account uncertainties, most data points are consistent with 0.8 in both models. The covering factor
variation from 0.76 to 0.9 is compatible with the compact object being deeply embedded into the stellar wind of the companion according to \citet{2021MNRAS.501.5892S}. From the long-term observations used here $C > 0.8$  at orbital phases [0.2-0.3] and eclipse.

\begin{table}%[scale=2.0,width=2\columnwidth]
 \caption{X-ray luminosity (10$^{37}$ erg s$^{-1}$) 
 and Comptonisation parameter}
    \centering
    \begin{tabular}{lccc} \toprule\toprule
      Orbital phase & L$_{x}$ (equation~(\ref{ec1})) & L$_{x}$ (equation~(\ref{ec4})) & $y = k\, T\, \tau^{2}/(m_{e}\, c^{2})$ \\ \midrule
Post-egress & 3$^{+14}_{-2}$ & 2.9$^{+1.8}_{-1.5}$ & $2.7\pm0.7$ \\
$[0.1-0.2]$ & 3$^{+6}_{-2}$ & 3$\pm$3 & $2.5\pm0.7$  \\
$[0.2-0.3]$ & 2.9$^{+1.6}_{-1.2}$ & 3$^{+2}_{-3}$ & 2.7$^{+0.7}_{-0.6}$ \\
$[0.3-0.4]$ & 3$^{+3}_{-2}$ & 3.0$^{+2.3}_{-1.7}$ & $1.8\pm0.8$ \\
$[0.4-0.5]$ & 3$^{+6}_{-2}$ & 3.0$^{+1.8}_{-2.1}$ & $2.4\pm0.7$ \\
$[0.5-0.6]$ & 3$^{+3}_{-2}$ & 2.6$\pm$1.8 & 2.3$\pm$0.6 \\
Pre-ingress & 2$^{+4}_{-1}$ & 2.3$^{+1.2}_{-0.9}$ & $1.6\pm0.6$ \\
Ingress & 1$^{+11}_{-1}$ & 1$^{+7}_{-1}$ & 3.2$^{+0.9}_{-0.8}$ \\
Eclipse & 0.10$^{+0.03}_{-0.02}$ & 0.10$\pm$0.03 & $3.7\pm0.7$ \\
Egress & 1.2$^{+0.3}_{-0.2}$ & 1.2$^{+0.3}_{-0.2}$ & $3.98\pm0.16$ \\
     \bottomrule
    \end{tabular}
    
    \label{tab:17}
\end{table}

\begin{figure}[hbtp]
  \centering
  \includegraphics[angle=-90.0,width=\columnwidth]{flujo_no_absorbido_compst.ps} 
  \includegraphics[angle=-90.0,width=\columnwidth]{profundidad_optica_compst.ps} 
  \includegraphics[angle=-90.0,width=\columnwidth]{norma_powerlaw_compst.ps}
  \includegraphics[angle=-90.0,width=\columnwidth]{compst_norm.ps} 
  \includegraphics[angle=-90.0,width=\columnwidth]{temperatura_compst.ps}
 \caption[]{\label{figure6} %
    Evolution of some parameters of the model described by equation~(\ref{ec4}). Unabsorbed flux is in units of 10$^{-9}$ erg cm$^{-2}$ s$^{-1}$. Power-law normalisation is expressed in units of keV$^{-1}$ s$^{-1}$ cm$^{-2}$. Filled red squares: initial values. Open blue circles: values obtained by constraining the range of values of the photon index (pre-ingress, ingress, eclipse and egress). 
}
\end{figure}

\begin{figure}[hbtp]
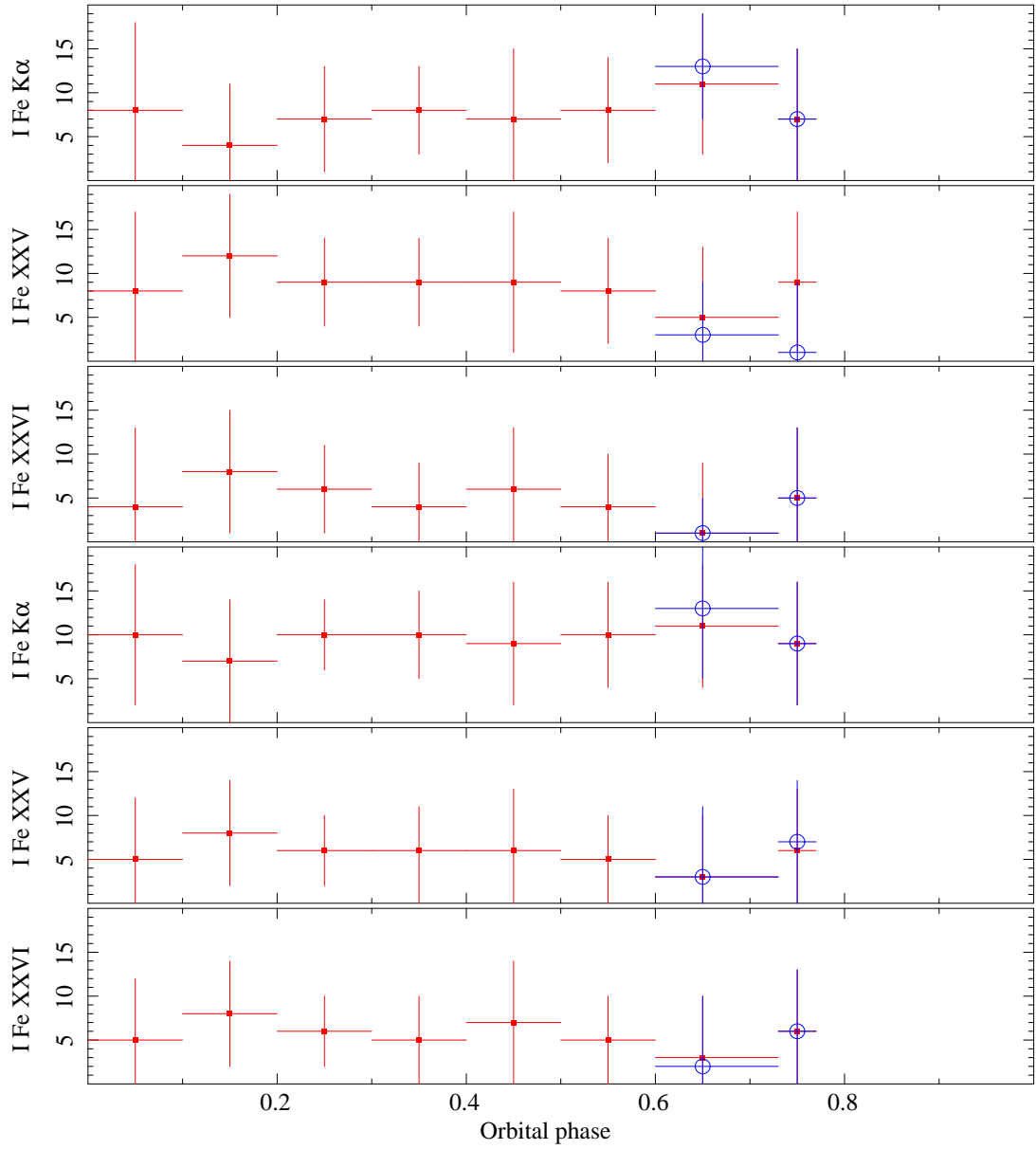

  \centering
  \includegraphics[angle=-90.0,width=\columnwidth]{intensidad_fe_k_alpha_bbody.ps} 
  \includegraphics[angle=-90.0,width=\columnwidth]{intensidades_bbody_hierro_25.ps}
  \includegraphics[angle=-90.0,width=\columnwidth]{intensidades_fe_26_bbody.ps}
  \includegraphics[angle=-90.0,width=\columnwidth]{intensidades_compst_fe_k_alpha.ps}
  \includegraphics[angle=-90.0,width=\columnwidth]{intensidades_fe_25_compst.ps}
  \includegraphics[angle=-90.0,width=\columnwidth]{intensidades_hierro_26_compst.ps} 
 \caption[]{\label{figure7} %
    Evolution of the Gaussian intensity of the Fe emission lines with orbital phase. \emph{Top, second and third panels:} model described by equation~(\ref{ec1}). \emph{Fourth, fifth and bottom panels:} model described by equation~(\ref{ec4}). The unit of the line flux $I$ is 10$^{-4}$ photons s$^{-1}$ cm$^{-2}$. Filled red squares: initial values. Open blue circles: values obtained by constraining the range of values of the photon index (pre-ingress, ingress, eclipse and egress).  
  }
\end{figure}

\begin{figure}[hbtp]
  \centering
  \includegraphics[angle=-90.0,width=\columnwidth]{eqw_bbody_fe_k_alpha.ps}
  \includegraphics[angle=-90.0,width=\columnwidth]{eqw_bbody_fe_25.ps}
  \includegraphics[angle=-90.0,width=\columnwidth]{eqw_bbody_fe_26.ps}
  \includegraphics[angle=-90.0,width=\columnwidth]{eqw_compst_fe_k_alpha.ps}
  \includegraphics[angle=-90.0,width=\columnwidth]{eqw_compst_fe_25.ps} 
  \includegraphics[angle=-90.0,width=\columnwidth]{eqw_compst_fe_26.ps} 
 \caption[]{\label{figure8} %
    Evolution of the equivalent width (EW) of the Fe emission lines with orbital phase. \emph{Top, second and third panels:} model described by equation~(\ref{ec1}). \emph{Fourth, fifth and bottom panels:} model described by equation~(\ref{ec4}). EW is in units of keV. Filled red squares: initial values. Open blue circles: values obtained by constraining the range of values of the photon index (pre-ingress, ingress, eclipse and egress). 
  }
\end{figure}

\begin{figure}[hbtp]
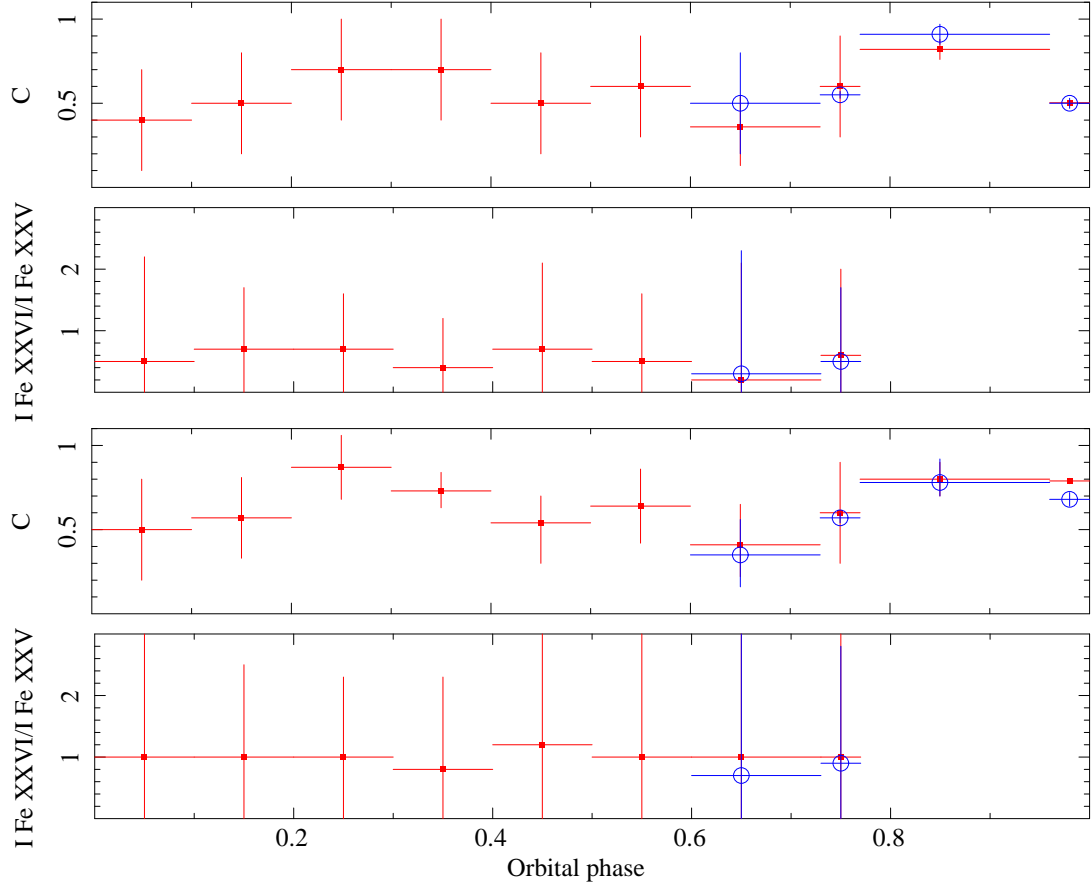

  \centering
  \includegraphics[angle=-90.0,width=\columnwidth]{covering_fraction_bbody.ps}
  \includegraphics[angle=-90.0,width=\columnwidth]{cociente_fe_26_fe_25_bbody.ps}
  \includegraphics[angle=-90.0,width=\columnwidth]{covering_fraction_compst.ps}
  \includegraphics[angle=-90.0,width=\columnwidth]{cociente_fe_26_fe_25_compst.ps}  
  \caption[]{\label{figure11} %
    Evolution of the partial covering fraction $C$ and line intensity ratio \ion{Fe}{xxvi}/\ion{Fe}{xxv}. \emph{Top and second panels}: model described by equation~(\ref{ec1}). \emph{Third and bottom panels}: model described by equation~(\ref{ec4}). Filled red squares: initial values. Open blue circles: values obtained by constraining the range of values of the photon index (pre-ingress, ingress, eclipse and egress).   
  }
\end{figure}

\subsection{Iron line complex}
\label{ironlines}

The spectral resolution of \maxi is not good enough to resolve the iron line complex, i.e. fluorescence emission lines (Fe K$\alpha$ at $\sim$6.4 keV, Fe K$\beta$ at $\sim$7.1 keV) and recombination emission lines (\ion{Fe}{xxv} at $\sim$6.7 keV and \ion{Fe}{xxvi} at $\sim$6.9 keV). From \maxi observations, the iron line feature is found to peak, in general, at $\sim$6.5 keV in the out-of-eclipse phase and peak at higher energies $\sim$6.7 keV in the ingress phase. This suggests the presence of both Fe K$\alpha$ and recombination lines. The high spectral resolution provided by other missions such as \asca, \xmm or \chandra, proved to be instrumental in resolving these emission lines, if present \citep{1996PASJ...48..425E,2019ApJS..243...29A,2021MNRAS.501.5892S}. Other sources where this dichotomy have been reported are Vela X$-$1 \citep{2014A&A...563A..70M,2013A&A...554A..37D,2016A&A...588A.100M}, 4U 1538$-$52 \citep{2011A&A...526A..64R,2015A&A...580A.140R} and GX 301$-$2 \citep{2011A&A...535A...9F,2014MNRAS.441.2539I}. Therefore, to describe the Fe complex, values obtained from \asca observations were adopted and used to fit our orbital phase-resolved spectra \citep{1996PASJ...48..425E}, where the energies and line widths were fixed and the line intensities were left as free parameters in the fit.

Unfortunately, none of the iron emission lines could be detected by \maxi neither in the eclipse nor in the egress phases (Figure~\ref{figure7}).

All free parameters, line intensities, equivalent widths and line intensity ratio \ion{Fe}{xxvi}/\ion{Fe}{xxv} (Figures~\ref{figure7}, ~\ref{figure8} and ~\ref{figure11}, respectively) were found consistent with constant values within errors out-of-eclipse.

\subsection{The stellar wind in \centau}
\label{averaged_spectrum_3}

Thanks to our orbital phase-resolved spectroscopy, we can study the properties of the stellar wind in \centau by analysing the variation of the hydrogen column $N^{1}_{H}$ along the orbit. As a first step, we have applied
a simple spherically stellar wind model to describe our observational data. This simple model reproduced the observed shape of the absorption curve along the entire orbit in the wind-fed HMXB 4U 1538-52 \citep{2008xru..confE..56R} %Rodes et al.
and it is consistent with the stellar wind description
of \centau reported by \citet{2001ApJ...547..973W}. %Wojdowski et al.

The radial flow velocity takes the form \citep{1975ApJ...195..157C,1982ApJ...259..282A}:

\begin{equation}\label{windvel} 
	v_w(r)=v_{\infty}\left( 1 - \frac{R_{c}}{r} \right)^{\alpha} \; , 
\end{equation} 

where $v_{\infty}$ is the terminal velocity of the wind, $R_{c}$ is the
radius of the companion star, $r$ is the distance from the
centre of the companion star and $\alpha$ is the velocity gradient.

\begin{figure}[htbp] 
  \centering
  \includegraphics[scale=0.10]{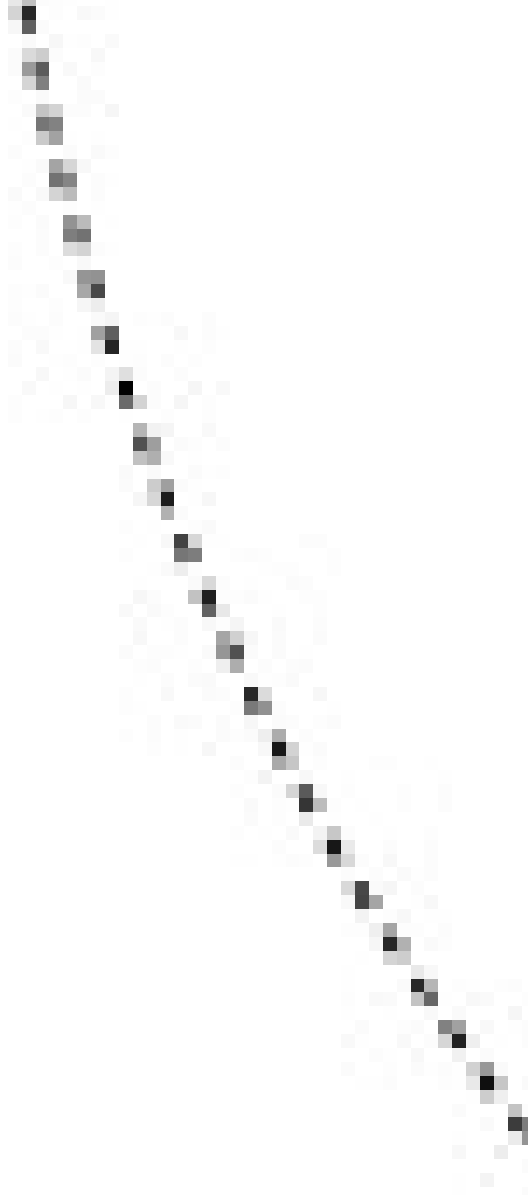} 
  \caption{Schematic view of the binary system. For simplicity we have assumed a circular orbit whose parameters $r$, $s$ and $\theta$ are shown. We note that in this sketch the orbital phase 0 (1) is centred on the mid-time of eclipse.}
\label{windesq} 
\end{figure}

Conservation of mass requires: 
\begin{equation}\label{masscons} 
	n_H = \frac{\dot{M}_c}{4 \; \pi \; r^2 \; v_w} \; , 
\end{equation} 
where $\dot{M}_c$ is the mass loss rate from the primary and $n_H$ is the wind density. Combining equations~(\ref{windvel}) and (\ref{masscons}) and integrating the wind density along the line of sight to the X-ray source, it is possible to find a model which describes the variation in $N_H$ with orbital phase properly. Defining $s$ as the distance through the stellar wind along the line from the compact object toward the observer (see Fig.~\ref{windesq}), we have: 
\begin{equation}\label{Nhequac} 
	N_H = \int^{s}_{0} n_H  \; ds = n_H \; s = n_H  \; 2 \; r \; 
	\sin \theta \; . 
\end{equation} 
The angle $\theta$ is related to the orbital phase and equation~(\ref{Nhequac}) can be rewritten by: 
\begin{equation}\label{Nhequac2} 
	N_H = \frac{\dot{M}_c}{4 \; \pi \; r^2 \; v_\infty \; \left( 
	1 - \frac{R_c}{r} \right)^\alpha} 
	 \; 2 \; r \;	\sin \left( \frac{\pi}{2} - 2 \; \pi \; \phi \right) \; . 
\end{equation} 

As we have mentioned before, $N^{1}_{H}$ represents the ISM plus the environment that surrounds the star. The stellar wind model is described by means of the following equation:

\begin{equation}
    N_{H} =\frac {X_{H}\, {\dot{M}}_{c}}{m_{H}\, 2\,\pi\, r\, v_{\infty} \left( 1 - \frac{R_{c}}{r} \right)^{\alpha}} \sin \left( \frac {\pi} {2} -2\, \pi\, \phi \right)
    \label{ec6}
\end{equation}

\begin{figure}[hbtp]
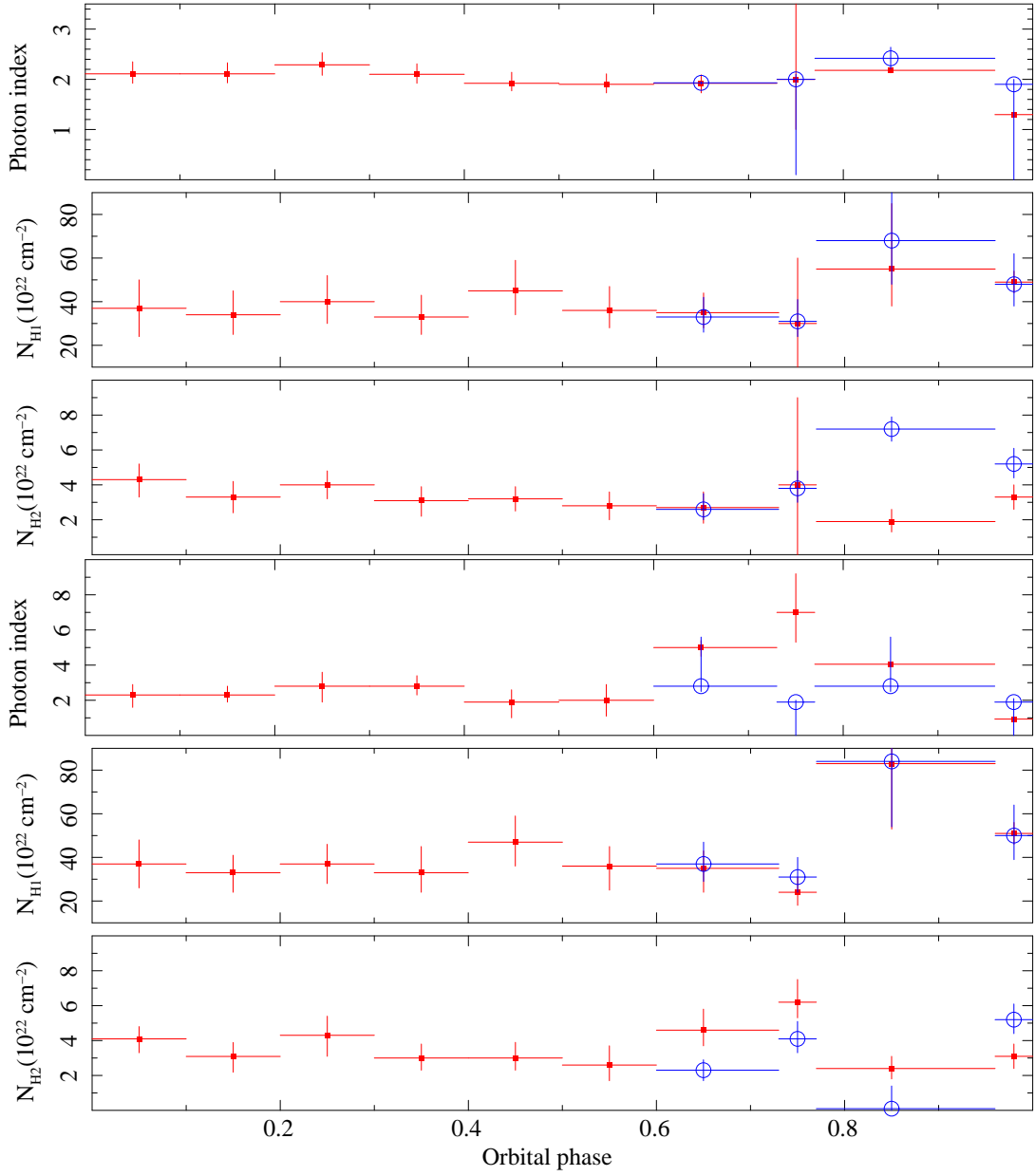

  \centering
  \includegraphics[angle=-90.0,width=\columnwidth]{photon_index_bbody_v2.ps}
  \includegraphics[angle=-90.0,width=\columnwidth]{columna_n_h_1_bbody.ps}
  \includegraphics[angle=-90.0,width=\columnwidth]{columna_n_h_2_bbody.ps}
  \includegraphics[angle=-90.0,width=\columnwidth]{photon_index_compst.ps}
  \includegraphics[angle=-90.0,width=\columnwidth]{columna_n_h_1_compst.ps}
  \includegraphics[angle=-90.0,width=\columnwidth]{columna_n_h_2_compst.ps}  
  \caption[]{\label{figure12} %
    Evolution of the photon index $\Gamma$ and absorption columns $N_{H}^1$ and  $N_{H}^2$. \emph{Top, second and third panels}: from equation~(\ref{ec1}). \emph{Fourth, fifth and bottom panels}: from equation~(\ref{ec4}). Filled red squares: initial values. Open blue circles: values obtained by constraining the range of values of the photon index (pre-ingress, ingress, eclipse and egress). 
  }
\end{figure}

\begin{figure}[hbtp]
  \centering
  \includegraphics[angle=-90.0,width=\columnwidth]{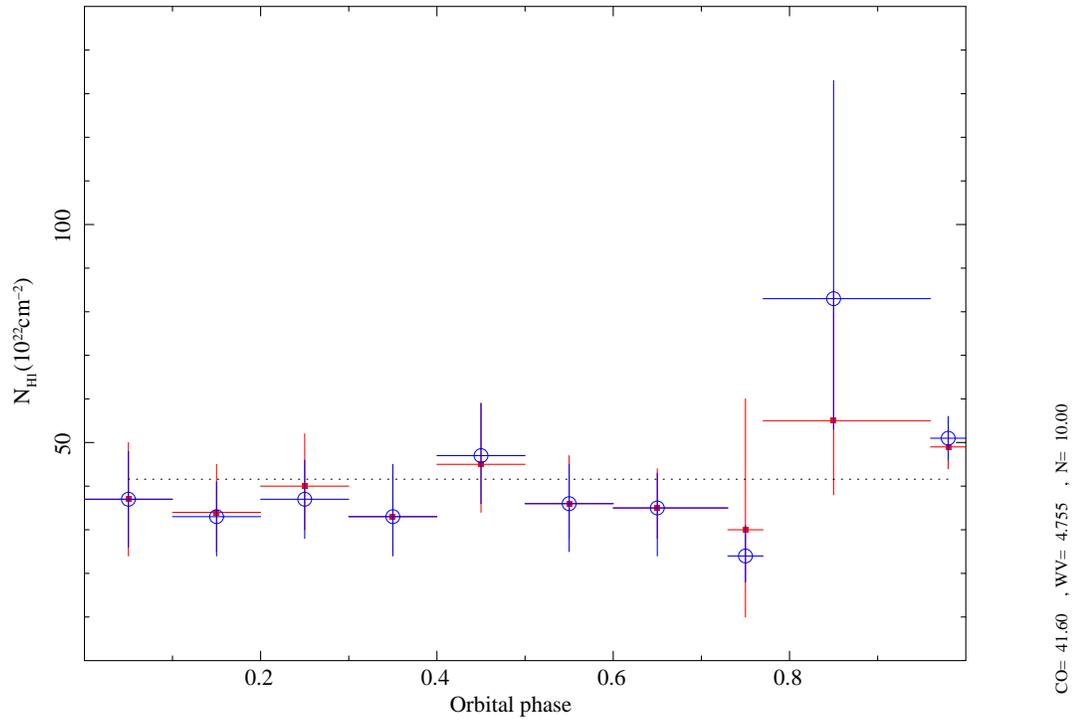} 
  \caption[]{\label{figure31} %
    Evolution of the absorption column $N^{1}_{H}$ (ISM plus local absorption). Red filled squares: from equation~(\ref{ec1}). Blue open circles: from equation~(\ref{ec4}). The dotted line represents the fit to a constant. A better fit could be obtained by applying equation~(\ref{ec6}), but the large uncertainties prevent us to draw firm conclusions. 
  }
\end{figure}

where $X_{H}$, the hydrogen mass fraction which is equal to 0.76, $m_{H}$ is the hydrogen atom mass, $\dot{M}_{c}$ the mass-loss rate, $v_{\infty}$ s the terminal velocity of the wind in the range [1\,000-3\,000] km\,s$^{-1}$ \citep{2015A&A...577A.130F} and $r$ represents the binary separation. Attempts were made to fit the local absorption $N^{1}_{H}$ using equation~(\ref{ec6}). A smooth wind model
could possibly describe the behaviour of our data, although the large uncertainties prevent us from drawing strong conclusions from this result.

The evolution of the absorption column for both models is shown in Figure~\ref{figure12}, and it is consistent with a constant value. The variation of the $N^{1}_{H}$ columns have been plotted together against the orbital phase in Figure~\ref{figure31}. As can be seen from the graph, both values are consistent within errors and the fit has been represented by a dotted line. \citet{2008ApJ...675.1487S} studied \centau over two consecutive orbits with observations taken by the RXTE observatory. They found that $N_{H}$ was increased near eclipse-ingress and egress during both orbits. However, during the second orbit, $N_{H}$ was relatively constant prior to mid-orbit and rose continuously afterward. They also interpreted the small increase in $N_{H}$ seen in the orbital phase [0.3--0.4] as bow shock in front of the accretion stream. The two-dimensional numerical simulations performed by \citet{1991ApJ...371..684B} provided the variation of the absorbing column of material, $N_{H}$, as a function of orbital phase). The dependence on binary separation showed a double-peaked structure, with one peak occurring slightly before orbital phase 0.5 and the second somewhat after (see their Figure 2). On the other hand, the orbital phase dependence of $N_H$ taking tidal stream and accretion wake into account showed small peaks associated with clumps in the wind and a strong jump at orbital phase 0.6 (see their Figure 8). They also plotted the column density of the undisturbed wind model showing a concave shape below the full simulated models. \citet{1990ApJ...356..591B} also showed in their simulations that the column density throughout the orbit changes between consecutive orbits due to variability in the accretion flow. We have not found this behaviour in our long-term \maxi observations although a small enhanced in $N^{1}_{H}$ could be present in the orbital phase [0.4--0.5] (see Figure~\ref{figure12}). However, it cannot be directly probed within the context of our \maxi analysis, which only allows inferring the long-term properties of the source averaged over several orbital phase bins. To sum up, the models carried out by \citet{1991ApJ...371..684B} do not well describe the observed behaviour for Cen X$-$3.

\section{Summary and conclusions}
\label{conclusion}

We have investigated the long-term variation of spectral parameters exploiting the continuous monitoring of \centau with \maxi and its outstanding spectral capabilities. From the analysis of the \maxi light curve, we estimated the orbital period of the binary system, $P_\mathrm{orb} = 2.0870\pm0.0005$ days, which agrees with the value given by \citet{1992ApJ...396..147N} and \citep{2017symm.conf..159R}, and we also noticed the presence of a superorbital period of $P_\mathrm{superorb} = 220\pm5$ days which is included in the interval [93.3-435.1] days reported by \citet{2014JPSCP...1a3104S} who detected it by using a power spectrum density technique.

We have described the X-ray spectra of \centau by two models consisting of a blackbody plus a power law and a Comptonisation of cool photons on hot electrons plus a power law, both modified by an absorption covering fraction factor and Gaussian functions, and performed detailed spectral (orbital phase-averaged and phase-resolved, in 2.0--20.0 keV) analysis. Our results can be summarised as follows:

-- The blackbody emitting area has an averaged radius of $0.71^{+0.19}_{-0.16}$ km and its size varies from 1 or 3 km out-of-eclipse, which is consistent with a hot spot on the NS surface, to $9\pm3$ km in eclipse covering a much larger area.

-- From the unabsorbed flux, the total X-ray luminosities in the 2.0--20.0 keV were found $(1.9^{+1.0}_{-0.8})\times 10^{37}$ \ergs (equation~(\ref{ec1})) and $(2^{+3}_{-1})\times 10^{37}$ \ergs (equation~(\ref{ec4})) on average. This high luminosity in the long term, one order of magnitude larger than that for wind-fed X-ray binaries, indicates that accretion must be enhanced by other mechanisms. In the case of \centau, this should be due to an accretion disc, matter co-rotating with the compact object and/or soft X-ray reflection from the inner accretion disc region. In addition, the out-of-eclipse X-ray luminosity was 10-30 times higher than in eclipse.

-- From the comparison of the high-state, low-state and averaged spectra, it has been deduced that the emission region is compatible in size and consistent with thermal emission produced at the NS polar cap. The values obtained for the high and low state luminosities suggest that the difference can be attributed to a drop in the accretion rate rather than an overall enhancement of absorption.

-- The column density of absorbing matter has two components: $N^{2}_{H} \sim (2-7)\times 10^{22}$ cm$^{-2}$ which represents the ISM towards the system and $N^{1}_{H} \sim (2-8)\times 10^{23}$ cm$^{-2}$ and both models yield similar column densities estimates within errors.

-- The spectra show the iron fluorescence line at $\sim$6.4 keV which is detected by \maxi but it cannot resolve recombination iron lines. Description of iron line complex adopted in this work suggest the presence of both near neutral iron line (Fe K$\alpha$) and highly ionised species (\ion{Fe}{xxv} and \ion{Fe}{xxvi}) except in the eclipse and egress spectra where none of the emission lines are detected. No modulation along the orbit is seen in the line intensities. 

-- The orbital dependence of the column density was compatible with a constant value within errors. The large value of the column density $N_H^1 \sim 4\times 10^{23}$ cm$^{-2}$ strongly favours a highly inhomogeneous surrounding environment. To constrain better \centau stellar wind properties it is necessary to improve uncertainties of the parameters, specially column density and EW of iron emission lines, for probing the geometry and distribution of circumstellar matter around the compact object. Numerical simulations performed by \citet{1991ApJ...371..684B} were not able to well describe the observed behaviour for Cen X$-$3.

\begin{acknowledgements}
  We thank the referee for the useful and constructive review which helped us to improve the content of this paper.
  This work was supported by the Spanish Ministry of Education and Science project number ESP2017-85691-P. This research made use of MAXI data provided by RIKEN, JAXA and MAXI team. JJRR acknowledges the support by the Matsumae International Foundation Research Fellowship No14G04, and also thanks the entire MAXI team for the collaboration and hospitality in RIKEN.  We would like to thank particularly T. Mihara and S. Nakahira for invaluable assistance in analysing MAXI data. This work has made use of data from the European Space Agency (ESA) mission Gaia (https://www.cosmos.esa.int/gaia), processed by the Gaia Data Processing and Analysis Consortium (DPAC, https://www.cosmos.esa.int/web/gaia/dpac). Funding for the DPAC has been provided by national institutions, in particular the institutions participating in the Gaia Multilateral Agreement.
\end{acknowledgements}

\bibliographystyle{rmaa} % style rmaa.bst
\bibliography{rm-journal-example} % your references jjrr_ads_v4_

\end{document}